\documentclass[12pt]{article}
\usepackage{amsmath}
\usepackage{amssymb}
\begin{document}
%\begin{flushright}
%BHU-PHYS-CAS Preprint\\
%arXiv: 1005.5067 [hep-th]
%\end{flushright}
\vskip 2cm
\begin{center}
{\bf {\Large   {Nilpotent Symmetries of a Model of 2D Diffeomorphism Invariant Theory: BRST Approach }}}

\vskip 3.0cm

{\sf R. P. Malik$^{(a,b)}$}\\
$^{(a)}$ {\it Physics Department, Institute of Science,}\\
{\it Banaras Hindu University, Varanasi - 221 005, (U.P.), India}\\

\vskip 0.1cm

%{\bf and}\\

\vskip 0.1cm

$^{(b)}$ {\it DST Centre for Interdisciplinary Mathematical Sciences,}\\
{\it Institute of Science, Banaras Hindu University, Varanasi - 221 005, India}\\
{\small {\sf {e-mail address: rpmalik1995@gmail.com}}}
\end{center}

\vskip 2cm
\noindent
{\bf Abstract:} Within the framework of Becchi-Rouet-Stora-Tyutin (BRST) formalism, we discuss
the {\it full} set of proper BRST and anti-BRST transformations for a 2D diffeomorphism invariant
theory which is described by the Lagrangian density of a standard  bosonic string. The above 
(anti-)BRST transformations are off-shell nilpotent and 
absolutely anticommuting. The latter property is valid on a submanifold of the space
of the quantum fields where the 2D version of the {\it universal} (anti-)BRST invariant 
Curci-Ferrari (CF) type of restrictions are satisfied. We derive the precise forms of the BRST 
and anti-BRST invariant Lagrangian densities as well as the exact expressions for the 
conserved (anti-)BRST and ghost charges. The lucid derivation of the 
{\it proper}  anti-BRST symmetry transformations and the emergence of the CF-type 
restrictions are completely  {\it novel} results for our present bosonic string which has
already been discussed {\it earlier} in literature where {\it only} the BRST symmetry transformations 
have been pointed out. We briefly mention the derivation of the CF-type restrictions from the {\it modified}
version of Bonora-Tonin superfield approach, too.\\

\vskip 2cm



\newpage

\section{Introduction}

One of the most exciting and captivating areas of research in theoretical high energy
physics (THEP), over the last few decades, has been the subject of (super)strings and related extended objects (see, 
e.g. [1-4] for details). This is due to the fact that, in one stroke, these theories
provide a possible scenario of unification of {\it all} the fundamental interactions of nature
and a promising candidate for the precise theory of quantum gravity. The modern developments
in the realm of (super)strings have influenced many other areas of research in THEP, e.g.
non-commutative field theories, higher $p$-form ($p = 2, 3, 4, ...$) gauge theories,
higher spin gauge theories, supersymmetric gauge theories and related mathematics, 
gauge-gravity duality, AdS/CFT correspondence, etc. The quantization of these
(super)string theories have led us to imagine about the higher dimensional view of the 
physical world we live in. It has been established that one cannot consistently quantize the 
dual-string theory [5] unless the spacetime dimension D = 26 and the intercept ($\alpha_0$) of the
leading Regge trajectory $\alpha_0 = 1$. These results have been obtained and formally 
established from many different considerations like the requirement of the validity of 
proper Lorentz algebra, unitarity requirements of these string theories, nilpotency 
of the Becchi-Rouet-Stora-Tyutin (BRST) charge, etc. In this context one of the earliest 
attempts to covariantly quantize a bosonic string theory, within the framework of BRST formalism, 
was undertaken by Kato and Ogawa [6] where the 2D diffeomorphism symmetry of this theory 
was exploited.

In the above work [6], it is precisely the infinitesimal version of the 2D {\it classical} diffeomorphism symmetry invariance
of the theory that has been primarily exploited to perform the BRST quantization where
{\it only} the BRST symmetries have been discussed. However, there is no discussion
about the anti-BRST symmetries and related Curci-Ferrari (CF)-type restrictions which are
the hallmarks of a properly BRST quantized theory. In this work [6], the inverse of the metric tensor
has been taken in such a manner that the conformal anomaly does {\it not} spoil the BRST analysis.
In fact, the inverse of the metric tensor has been decomposed such that it has {\it three} independent  degrees 
of freedom to begin with. A Lagrange multiplier  field density has been incorporated into the 2D Lagrangian density so that the equation of
motion w.r.t. {\it it} puts a restriction on the determinant of the {\it above} metric tensor. The latter condition
reduces the independent degrees of freedom of the metric tensor from {\it three} to {\it two}. 
The BRST charge has been calculated in the flat limit where the metric tensor becomes 
Minkowskian in nature (see, e.g. [6] for more details). The nilpotency requirement of this
BRST charge leads to the derivation of D = 26 and $\alpha_0 = 1$. One of the central theme 
of our present investigation is to focus on the existence of (i) the proper anti-BRST symmetries
(corresponding to the BRST transformations taken in [6]),  and (ii) the (anti-)BRST invariant CF-type
of restrictions which are responsible for the absolute anticommutativity of the nilpotent (anti-)BRST symmetry
transformations. We have taken, a very modest step in this direction, in our present endeavor.

We have performed the full BRST analysis of the above 
theory [6] in the sense that we have derived the {\it proper} anti-BRST
symmetry transformations corresponding to the BRST symmetry transformations that have been taken 
into account in [6]. The BRST and anti-BRST symmetry transformations are found to be
off-shell nilpotent and absolutely anticommuting in nature. The latter property has been  
shown to be  true on a submanifold of the quantum Hilbert space of quantum fields where the CF-type restrictions (11) (see below) 
are satisfied. We 
observe that these restrictions are BRST as well as anti-BRST invariant thereby implying 
that these are {\it physical} restrictions (which can be imposed from {\it outside} on our 
present theory). We have derived, in our present endeavor, the BRST and anti-BRST invariant
Lagrangian densities and have shown explicitly their BRST and anti-BRST invariance. The conserved
charges of the theory have been computed in the flat limit where $A_0 = A_1 = 0, A_2 = 1$ [cf. Eq. (3) and related footnote]. In fact,
the latter conditions imply that the metric tensor of the theory transforms as:
${\tilde g}^{ab} \to \eta^{ab}$ where $\eta^{ab}$ is the flat metric of the 2D Minkowski space (which
is nothing but the 2D surface traced out by the propagation of the bosonic string). We have
also established that the standard algebra between the ghost charge and BRST charge (as well as
between the ghost charge and anti-BRST charge) is satisfied. We have also commented, very briefly,
about the nilpotency properties of the BRST and anti-BRST charges which are true at the {\it quantum} level
only when D = 26 and $\alpha_0 = 1$ provided we take into account the normal mode expansions of the
fields (consistent  with the appropriate boundary conditions) and substitute them in the computation and proof of nilpotency:
$Q_B^2 = \frac{1}{2}\, \{Q_B, \, Q_B\} = 0$ and $\bar Q_B^2 = \frac{1}{2}\, \{\bar Q_B, \, \bar Q_B\} = 0$.

The main motivating factors behind our present investigation are as follows. First, in the BRST
description [6] of the present bosonic string, {\it only} the BRST transformations have been
discussed corresponding to the infinitesimal diffeomorphism symmetry transformation of the theory. The nilpotent anti-BRST 
symmetry transformations have remained untouched in [6]. Thus, it is important for us to discuss the BRST 
as well as anti-BRST symmetry transformations {\it together} for the {\it complete} BRST analysis of  our 
present theory. We have accomplished this goal in our present endeavor. Second, in the BRST description of 
Kato and Ogawa [6], the auxiliary fields have been modified/redefined in a very complicated fashion to simplify the
theoretical analysis of the present theory. There is, however,  no basic physical arguments to support such kind
of modifications/redefinitions. We have, in our present endeavor, {\it not} invoked any such kind of 
modifications/redefinitions as our analysis is very straightforward. Third, the hallmark of a quantum theory, discussed
within the framework of BRST formalism, is the existence of the (non-)trivial Curci-Ferrari (CF) type
restrictions. We have derived such restrictions in our present endeavor which ensure the absolute 
anticommutativity of the (anti-)BRST symmetry transformations. Finally, our present work is important
because, for this model, the recently developed superfield approach [7] would be very useful as
our theory is diffeomorphism invariant. We hope that the application of  this superfield formalism 
[7] would shed some {\it new} lights on some specific aspects of our present theory (as far as the  
symmetries are concerned). In our Appendix D, we have briefly discussed about the applications of {\it this} superfield
approach which has been christened by us as the {\it modified} version of Bonora-Tonin  superfield approach (MBTSA).

Our present paper is organized as follows. To set up the notations and convention, we discuss very briefly
the diffeomorphism symmetry as well as the corresponding BRST approach in Sec. 2 which has been performed in
[6]. Our Sec. 3 is devoted to the discussion of BRST and anti-BRST symmetries where we also point out the
existence of the 2D version of the {\it universal} CF-type restrictions. We prove the (anti-)BRST invariance 
of this restriction and 
we demonstrate the nilpotency as well as the absolute anticommutativity of the (anti-)BRST symmetry 
transformations. We derive the explicit form of the BRST as well as anti-BRST invariant Lagrangian 
densities in Sec. 4. The conserved charges, corresponding to the continuous {\it internal} symmetries 
of the theory, are derived in Sec. 5, in the {\it flat} limit. Finally, we make some concluding remarks
on our present investigation in Sec. 6 and point out a few future directions for further investigations.

In our Appendices A, B and C, we incorporate some of the algebraic expressions as well as 
equations that have been used in the main body of our text. In our Appendix D, we concisely
discuss the derivation of the CF-type restrictions by using the {\it modified} version of the Bonora-Tonin
superfield approach (MBTSA) to BRST formalism.

\section{Preliminary: Diffeomorphism and BRST Invariance}

We begin with the Lagrangian density of a bosonic string theory as (see, e.g. [6] for details) 
\begin{eqnarray}
{\cal L}_0 = - \frac{1}{2k}\, {\tilde g}^{ab}\,\partial_a \,X^\mu \, \partial_b \, X_\mu 
+ E\, (\mbox{det} \,\tilde g + 1),
\end{eqnarray}
where ${\tilde g}^{ab} = \sqrt{- g} \,g^{ab}$ has two independent degrees of freedom\footnote{The
original Lagrangian density: $- \frac{1}{2k}\, \sqrt{- g} \,g^{ab}\,\partial_a \,X^\mu \, \partial_b \, X_\mu $ (with $k$ as the string tension parameter)
is endowed with the local conformal invariance. However, this conformal invariance
is broken by the conformal anomaly [8,9] if we regularize the system in a gauge-invariant 
manner. We have avoided this problem by taking ${\tilde g}^{ab} = \sqrt{- g} \,g^{ab}$ as
the metric tensor of our present theory [6] with {\it three} independent degrees of freedom to start with. The EoM w.r.t. $E$
(i.e. $\mbox{det} \,{\tilde g} = - 1$) reduces the independent degrees of freedom  of the above specifically defined metric tensor from 
{\it three} to {\it two}.} because 
$\mbox{det} \,{\tilde g} = - 1 $ due to the equation of motion w.r.t. the Lagrange multiplier 
field $E$ which happens to be a scalar density [cf. Eq. (2) below]. Here the 2D surface, traced 
out by the propagation of the bosonic string, is parameterized by 
$\xi^a = (\xi^0, \,\xi^1) \equiv (\tau,\, \sigma)$ where $a = 0, 1$ and component parameters ($\tau, \, \sigma$) satisfy: 
$- \, \infty  <  \tau < +\, \infty $ and $0\, \leq \sigma \leq \pi $. The string coordinates 
$X^\mu (\xi)$ (with $\mu = 0, 1, 2, 3, ..., D-1$) are in the D-dimensional flat Minkowskian 
spacetime manifold and  ${\tilde g}^{ab} = \sqrt{- g} \,g^{ab}$ is the metric  tensor constructed 
with the determinant ($g = \mbox{det}\, g_{ab}$) and inverse ($g^{ab}$) of the metric tensor 
$g_{ab}$ of the 2D parameter space. Under the infinitesimal diffeomorphism
transformations\footnote{It will be noted that we differ from [6] by an overall
sign factor in the diffeomorphism transformations (2) and BRST transformations (6) because we have chosen
the infinitesimal diffeomorphism transformation $\xi^a \to \xi^a - \varepsilon^a (\xi)$ whereas 
the same transformation has been taken as: $\xi^a \to \xi^a +\; \varepsilon^a (\xi)$ in [6].}
: $\xi^a \to \xi^a - \varepsilon^a (\xi)$, we have the following transformations 
$(\delta_\varepsilon)$ on the relevant fields of our present bosonic string  
theory\footnote{We choose the Latin indices $a, b, c, ..., l, m, n,... = 0, 1$
to denote $\tau$ and $\sigma$  directions on  the 2D surface (traced out by 
the propagation of the bosonic string) and the Greek
indices $\mu, \nu, \lambda, ... = 0, 1, 2, ...,  D-1$ stand for the spacetime directions of the 
D-dimensional flat Minkowskian spacetime manifold corresponding to the target space. The 
above 2D surface is embedded in the D-dimensional Minkowskian flat target space (which turns out to be 
26 at the {\it quantum} level). Throughout the whole body of our text, we denote the BRST and anti-BRST 
symmetry transformations by the symbols $s_B$ and $\bar s_B$, respectively. We adopt the convention of 
left-derivative w.r.t. the fermionic fields ($C^a, \bar C^a$, etc.) of our present theory. Consistent
with this convention, the Noether conserved currents in equations (25) and (26) are defined (see below).}, namely;
\begin{eqnarray}
&& \delta_\varepsilon \, X^\mu  = \varepsilon^a \,\partial_a X^\mu , \qquad 
\delta_\varepsilon \, E = \partial_a (\varepsilon^a \, E), 
\qquad \delta_\varepsilon \,(\mbox{det} \,{\tilde g}) = \varepsilon^a \,\partial_a (\mbox{det}\, {\tilde g} ),\nonumber\\
&& \delta_\varepsilon \,{\tilde g}^{ab} = \partial_m (\varepsilon^m \, {\tilde g}^{ab}) \, - \, 
(\partial_m \varepsilon^a) \,{\tilde g}^{mb} \, - \, (\partial_m \, \varepsilon^b)\, {\tilde g}^{am}, 
\end{eqnarray}
where $\varepsilon^a (\xi$) are the infinitesimal diffeomorphism transformation parameters. The above transformations 
leave the Lagrangian density (1) quasi-invariant [i.e. $\delta_\varepsilon \, {\cal L}_0 = 
\partial_a \, (\varepsilon^a \,{\cal L}_0)$]. This demonstrates that the action integral 
$S = \int d^2 \xi \, {\cal L}_0 \equiv \int_{- \, \infty}^{+ \, \infty} d\tau \int_{0}^{\pi} d\sigma\, {\cal L}_0$
remains invariant under the diffeomorphism transformations (2) provided the boundary conditions:
$\varepsilon^a (\xi) = 0$ at $\sigma = 0$ and $\sigma = \pi$ are imposed on the diffeomorphism parameter
$\varepsilon^a (\xi)$.

For the BRST quantization of the Lagrangian density (1), we have to invoke the gauge-fixing 
conditions. This can be achieved if we take the following decomposition for the metric tensor 
${\tilde g}^{ab}$ (see, e.g. [6] for details) 
\begin{eqnarray}
{\tilde g}^{ab} = 
\begin{pmatrix} 
A_1 + A_2  & A_0  \\ A_0 & A_1 - A_2 
\end{pmatrix},
\end{eqnarray}
and set the gauge-fixing conditions $A_0 = A_1 = 0$ so that we obtain $\det \,{\tilde g} = - 
{A_2}^2 = -\, 1.$ This shows that, for the choice $A_2 = 1$, we obtain the flatness condition\footnote{We
shall take the {\it flatness condition} ${\tilde g}^{ab} \to \eta^{ab}$ in the language of 
restrictions on the component gauge fields: $A_0 = A_1 = 0, \, A_2 = 1$ for the full discussion of our
theory within the framework of BRST formalism.} ${\tilde g}^{ab} \to \eta^{ab}$ with the signatures (+ 1, -  1).
By exploiting the standard techniques of the BRST formalism [10,11], we obtain the gauge-fixing and Faddeev-Poppov ghost
terms for the theory, in the language of the nilpotent ($s_B^2 = 0$) BRST transformations $s_B$, as 
(see, e.g. [10,11] for details)
\begin{eqnarray}
{\cal L}_{GF} + {\cal L}_{FP} = s_B\,\big[-\,i\,\bar C_0 A_0 - i\,\bar C_1 A_1 \big],
\end{eqnarray}
where $\bar C_0$ and $\bar C_1$ are the anti-ghost fields with ghost number (-1). 
It will be noted that the transformations $s_B \bar C_0 = i\, B_0$ and $s_B \bar C_1 = i\, B_1$ lead to
the emergence of the Nakanishi-Lautrup type auxiliary fields of the theory as $B_0$ and $B_1$ and the nilpotency
requirements produce $s_B B_0 = s_B B_1 = 0$. A close look at the transformations (2) and decomposition (3) leads to
the following BRST symmetry transformations for the component gauge fields
\begin{eqnarray}
&&s_B A_0 = C^a\,\partial_a\, A_0 - (\partial_0 \, C^1 + \partial_1 \, C^0)\,  A_1 - 
(\partial_0 \, C^1 - \partial_1 \, C^0) \, A_2, \\ \nonumber
&&s_B A_1 = C^a\,\partial_a \, A_1 - (\partial_0 \, C^0 - \partial_1 \, C^1) \, A_2 - 
(\partial_1 \, C^0 + \partial_0 \,  C^1)\, A_0, \\ \nonumber
&&s_B A_2 = C^a\,\partial_a \, A_2 - (\partial_0 \, C^0 - \partial_1 \, C^1) A_1 - 
(\partial_1 \, C^0 - \partial_0 \, C^1) A_0, 
\end{eqnarray}
where we have taken the replacement $(\varepsilon^a\longrightarrow C^a)$ which implies that the 
infinitesimal diffeomorphism parameters ($\varepsilon^a, \, a = 0, 1$) have been replaced 
by the fermionic $[(C^a)^2 = 0, \, C^a\, C^b + C^b \, C^a = 0]$ ghost fields $C^a$. As a 
consequence of this replacement, we have the following BRST symmetry transformations 
{\it vis-\`a-vis} the transformations (2), namely;
\begin{eqnarray}
&&s_B X^\mu = C^a \, \partial_a \, X^\mu, \qquad  s_B\, E = \partial_a \,(C^a E),\qquad  
s_B(\mbox{det} \, {\tilde g}) = \varepsilon^a \, \partial_a \, (\mbox{det} {\tilde g}), \nonumber\\
&&s_B C^a = C^b \, \partial_b \, C^a,\,\,\qquad s_B \bar C^a = i B^a,\qquad  s_B \,B^a = 0,\nonumber\\
&& s_B\,{\tilde g}^{ab} = \partial_m (C^m \, {\tilde g}^{ab}) \, - \, 
(\partial_m\,C^a) \,{\tilde g}^{mb} \, - \, (\partial_m \, C^b)\, {\tilde g}^{am}, 
\end{eqnarray}
where the transformation $s_B \,C^a = C^b \, \partial_b \, C^a$ has been derived from the 
requirement of the  nilpotency condition ($s_B^2 \, X^\mu = 0$).  With the inputs from (5) and (6), 
we obtain the BRST invariant Lagrangian density (${\cal L}_B$) from (4) and (1), modulo some
total derivatives, as\footnote{The Lagrangian density ${\cal L}_B$ has been written, modulo 
some total derivatives, in such a manner that BRST transformations (5) could be implemented 
in a simple and straightforward manner.}: 
\begin{eqnarray}
{\cal L}_B &=& {\cal L}_0 + B_0 A_0 + B_1 A_1 + i\, \big[C^a\partial_a \bar C_0 - \bar C_0 (\partial_a C^a)
- \bar C_1 (\partial_0 C^1 + \partial_1 C^0)\big]\, A_0 \nonumber\\
&+& i \,\big[ C^a \,\partial_a \,\bar C_1 - \bar C_1 \,(\partial_a \, C^a) -
 \bar C_0 (\partial_0 C^1 + \partial_1 C^0)\big] \,A_1 \nonumber\\ 
&-& i \,\big[\bar C_0 \,(\partial_0 \, C^1\, - \partial_1 \, C^0) + \bar C_1\, (\partial_0 C^0\, - \partial_1 C^1)\big]\, A_2. 
\end{eqnarray}
The above {\it full} Lagrangian density, under the {\it flatness} limit $ A_0 = A_1 = 0, A_2 = 1$, 
reduces to the following Lagrangian density\footnote{We would like to point out that the 
flatness limit (i.e. $ A_0 = A_1 = 0, A_2 = 1$) has been taken in all the terms of (7) except 
the gauge-fixing terms (i.e. $B_0\,A_0 + B_1\,A_1$) and the Lagrange multiplier term (i.e. $E\, (1 - A_2^2)$)
because the EoM w.r.t. $B_0, \, B_1$ and $E$ imply the same thing (i.e. $ A_0 = A_1 = 0, A_2 = 1$).}
\begin{eqnarray}
{\cal L}_B \longrightarrow {\cal L}_B^{(0)} &= &-\,\frac{1}{2\kappa }\,\eta^{ab}\partial_a X^\mu \partial_b X_\mu
+ E\,(1-A_2^2) + B_0\, A_0 + B_1\, A_1 \nonumber\\    &-& i\,\big[\bar C_0 \, (\partial_0 \, C^1 - \partial_1 \, C^0) +
\bar C_1\, (\partial_0 \,C^0 -\, \partial_1 \,C^1) \big],
\end{eqnarray}
which has been obtained in [6] after taking the help of the redefinitions of the auxiliary fields in a 
complicated fashion. In fact, these redefinitions are mathematical in nature and there is {\it no} physical
arguments to support the specific choices that have been made in [6] for the 
simplification of the Lagrangian  density in the {\it flat} space. We have obtained (8) from (7) 
in a straightforward manner (without any redefinitions/modifications, etc.).

\section{BRST and Anti-BRST Symmetries: Key Features}

It can be checked, in a straightforward fashion, that the BRST symmetry transformations, 
quoted in (5) and (6), are nilpotent of order two
%\footnote{It is rather algebraically more 
%involved to prove the nilpotency property (i.e. $s_{ab}^2 = 0$) in the case of the metric tensor 
%(i.e. $s_{ab}^2 \,{\tilde g}^{ab} = 0$). However, this statement about the sanctity of nilpotency is true.} 
(i.e. $s_B^2 = 0$). The proper anti-BRST
symmetry transformations, corresponding to the BRST transformations (6), are
\begin{eqnarray}
&&\bar s_B\, X^\mu = \bar C^a \, \partial_a X^\mu, \quad \qquad \bar s_B \,\bar C^a = \bar C^b \, \partial_b \,\bar C^a, \qquad 
\quad \bar s_B \, C^a = i\, \bar B^a, \nonumber\\
&& \bar s_B \,E = \partial_a \, (\bar C^a \,E), \qquad  \bar s_B \, (\mbox{det} {\tilde g}) = 
\bar C^a \, \partial_a (\mbox{det} {\tilde g}), \qquad \bar s_B\, \bar B^a = 0,\nonumber\\
&&\bar s_B\, {\tilde g}^{ab} = \partial_m \,(\bar C^m\, {\tilde g}^{ab}) - (\partial_m \,\bar C^a) \,{\tilde g}^{mb}
-  (\partial_m \,\bar C^b) \,{\tilde g}^{am},
\end{eqnarray}
which are off-shell nilpotent ($\bar s_B^2 = 0$) of order two. It will be  noted that we have invoked a new
Nakanishi-Lautrup type of auxiliary field $\bar B^a (\xi)$ in our theory. Thus, we observe that the
symmetry transformations (9), (6) and (5) satisfy {\it one} (i.e. nilpotency) of the two sacrosanct properties 
(i.e. nilpotency and absolute anticommutativity) that have to be satisfied by  any {\it proper} 
(anti-)BRST symmetry transformations. We further note that the last entry of (9) can be written 
in terms of $A_0, \, A_1, \, A_2$ in the following form, namely;
\begin{eqnarray}
&&\bar s_B A_0 = \bar C^a\,\partial_a\, A_0 - (\partial_0 \, \bar C^1 + \partial_1 \, \bar C^0)\,  A_1 - 
(\partial_0 \, \bar C^1 - \partial_1 \,\bar C^0) \, A_2, \nonumber\\
&&\bar s_B A_1 = \bar C^a\,\partial_a \, A_1 - (\partial_0 \, \bar C^0 - \partial_1 \, \bar C^1) \, A_2 - 
(\partial_1 \,\bar C^0 + \partial_0 \, \bar C^1)\, A_0,  \nonumber\\
&&\bar s_B A_2 = \bar C^a\,\partial_a \, A_2 - (\partial_0 \, \bar C^0 - \partial_1 \, \bar C^1) A_1 - 
(\partial_1 \, \bar C^0 - \partial_0 \, \bar C^1) A_0.
\end{eqnarray}
 Thus, it is clear that the anti-BRST transformations for $A_0, \, A_1, \, A_2$ are {\it exactly} same as 
 equation (5) with the replacement: $C^a \to \bar C^a$.

We dwell a bit now on the absolute anticommutativity property (i.e. $\{s_B, \bar s_B\} = 0$) of 
the (anti-)BRST symmetry transformations (10), (9), (6) and (5). It turns out that the requirement of 
$\{s_B, \, \bar s_B\} \, X^\mu = 0$ leads to the existence of the following Curci-Ferrari (CF)-type restrictions
(which are primarily {\it two} in numbers), namely;
\begin{eqnarray}
B^a + \bar B^a + i\, \big(C^b \, \partial_b \, \bar C^a + \bar C^b \, \partial_b \, C^a \big) = 0,
\qquad (a, b = 0, 1).
\end{eqnarray}
It turns out that the above conditions (11) have to be imposed to obtain the absolute anticommutativity 
(i.e. $\{s_B, \bar s_B\} = 0$) property when {\it all} the relevant fields of the whole theory are
taken into account. For instance, it can be checked that the requirement of 
$\{s_B, \, \bar s_B\} \, E = 0$ {\it also} requires the validity of the CF-type restrictions (11).  
Furthermore, we obtain the following (anti-)BRST symmetry transformations on the Nakanishi-Lautrup 
auxiliary fields $B^a (\xi)$ and $\bar B^a (\xi)$ due to the requirement of the absolute 
anticommutativity property (e.g. $\{s_B, \, \bar s_B\} \, C^a = 0$ and  
$\{s_B, \, \bar s_B\} \, \bar C^a = 0$), namely;
\begin{eqnarray}
s_B \, \bar B^a = C^b \, \partial_b \bar B^a - \bar B^b\,\partial_b\,C^a, \qquad 
\bar s_B\, B^a = \bar C^b \, \partial_b \, B^a -  B^b \,\partial_b \,\bar C^a.
\end{eqnarray}
Interestingly, the above transformations {\it also} satisfy the off-shell nilpotency 
property (i.e. $s^2_B = 0, \, \bar s^2_B = 0$) 
which is one of the key requirements of a proper set of (anti-)BRST symmetry transformations. Thus, we 
note that the (anti-)BRST symmetry transformations (12), (10), (9), (6) and (5) satisfy the
off-shell nilpotency ($s_B^2 = \bar s_B^2 = 0$)
and absolute anticommutativity ($\{ s_B, \bar s_B \} = 0 $) 
on a submanifold in the 2D Hilbert space of quantum fields
where the CF-restrictions (11) are satisfied.

We would enumerate here some of the subtle features associated with the CF-type restrictions (11) which 
are at the heart of the absolute anticommutativity property of our BRST and anti-BRST symmetry 
transformations. We note that these 2D restrictions on the auxiliary and (anti-)ghost fields are (anti-)BRST invariant quantity, namely; 
\begin{eqnarray}
&&\bar s_B\, \Big[B^a + \bar B^a + i\, \Big(C^b \, \partial_b \, \bar C^a 
+ \bar C^b \, \partial_b \, C^a \Big)\Big] = 0, \nonumber\\
&&s_B\, \Big[B^a + \bar B^a + i\, \Big(C^b \, \partial_b \, \bar C^a 
+ \bar C^b \, \partial_b \, C^a \Big)\Big] = 0.
\end{eqnarray}
This demonstrates that the CF-type restrictions of our present theory are {\it physical} 
(in some sense) and the {\it submanifold} of the quantum fields in the Hilbert space defined by {\it it} is physically relevant. 
This demonstrates that
our (anti-)BRST invariant theory is {\it consistently} defined on the submanifold where the 
CF-type restrictions (11) are always valid. In fact, on {\it this} submanifold {\it alone}, 
the BRST and anti-BRST symmetry transformations have their own identities as they are 
linearly independent of each-other (due to their absolute  anticommutativity). In the proof 
of (13), it is obvious that we have taken into account the (anti-)BRST transformations (12), (9) 
and (6) as well as the above {\it specific} submanifold.

We end this section with the following remarks on the nilpotency properties (i.e. $s^2_B = 0, \, \bar s^2_B = 0$) 
associated with the BRST and anti-BRST symmetry transformations  $s_B $ and $\bar s_B$. 
First of all, we note that $s^2_B \, X^\mu = 0$ leads to the derivation of
$s_B \, C^a = C^b \,\partial_b \, C^a$ where $s^2_B \, C^a = 0$ is {\it also} satisfied. In exactly similar
fashion, we obtain $\bar s_B \, \bar C^a = \bar C^b \,\partial_b \, \bar C^a$ (where $\bar s^2_B \, \bar C^a = 0$)
from the requirement of nilpotency of the anti-BRST symmetry transformation on $X^\mu$ field 
($\bar s^2_B \, X^\mu = 0$). The proof of the nilpotency (i.e. $s^2_B = 0, \, \bar s^2_B = 0$) 
of the transformations $s_B\, {\tilde g}^{ab}$ and $\bar s_B\, {\tilde g}^{ab}$ [cf. Eqs. (6) and (9)] is 
algebraically more involved\footnote{We have collected some of the crucial expressions (as well as
equations) in our Appendix A which establish the nilpotency ($s^2_B\, {\tilde g}^{ab} = 0$) of the BRST transformations when
they act on the metric tensor ${\tilde g}^{ab}$.}.
 However, it turns out that  $s^2_B \,{\tilde g}^{ab} = 0$ and 
$\bar s^2_B \,{\tilde g}^{ab} = 0$ are {\it indeed} true. This proof, in turn, implies that the (anti-)BRST 
transformations [cf. Eqs. (5) and (10)] of the component gauge fields (i.e. $A_0, \, A_1, A_2$) 
are {\it automatically} nilpotent (i.e. $s^2_B = 0, \, \bar s^2_B = 0$) of order two.

\section{(Anti-)BRST Invariant Lagrangian Densities}

We have already mentioned the BRST invariant Lagrangian densities (7) and (8) in the (non-)flat limits. In
our present section, we establish their BRST invariance. The analogue of the Lagrangian density (7) that
remains invariant, under the anti-BRST symmetry transformations (9), (10) and (12), is as follows [10,11]:
\begin{eqnarray}
{\cal L}_{\bar B} = {\cal L}_0 \, + \, {\bar s}_B \, \big[i\,C_0 \, A_0 + i\,C_1 \, A_1 \big].
\end{eqnarray}
Using the explicit anti-BRST symmetry transformations (9), (10) and (12), we obtain the following
explicit form of the anti-BRST invariant Lagrangian density ${\cal L}_{\bar B}$ as
\begin{eqnarray}
{\cal L}_{\bar B} & = &{\cal L}_0 - {\bar B}_0 A_0 - \bar B_1 A_1 
+ i\,\Bigl [C_0 (\partial_a \, \bar C^a) + (\partial_a C_0)\, \bar C^a 
+ C_1 \, (\partial_0 \, \bar C^1 + \partial_1 \, \bar C^0)\Bigr] \, A_0 \nonumber\\
&+& i\,\Bigl [C_1 \, (\partial_a \, \bar C^a) + (\partial_a \, C_1) \,\bar C^a 
+ C_0 \, (\partial_0 \, \bar C^1 + \partial_1 \, \bar C^0)\Bigr] \, A_1  \nonumber\\
& + & i\,\Bigl[C_0 \, (\partial_0 \, \bar C^1 - \partial_1 \, \bar C^0) 
+ C_1 \,(\partial_0 \, \bar C^0 - \partial_1 \, \bar C^1) \Bigr] \, A_2,
\end{eqnarray}
where some total derivative terms have been dropped as they do not affect the dynamics of the theory. 
The flat limit (i.e. $A_0 = A_1 = 0, \, A_2 = 1$) of the above Lagrangian density, 
in its full blaze of glory, is as follows:
\begin{eqnarray}
{\cal L}_{\bar B} \to {\cal L}_{\bar B}^{(0)} &=& -\,\frac{1}{2\kappa }\,\eta^{ab}\partial_a X^\mu \partial_b X_\mu
+ E\,(1-A_2^2) - \bar B_0\, A_0 - \bar B_1\, A_1 \nonumber\\   
& + & i\,\bigl[C_0 \, (\partial_0 \, \bar C^1 - \partial_1 \, \bar C^0) 
+ C_1 \,(\partial_0 \, \bar C^0 - \partial_1 \, \bar C^1) \bigr],
\end{eqnarray}
where the above limit has {\it not} been imposed on the gauge-fixing terms ($- \bar B_0\, A_0 - \bar B_1\, A_1 $)
and the term $(E\,(1-A_2^2))$ with the Lagrange multiplier field. 
We shall be calculating the conserved charges of the theory from the Lagrangian densities 
(8) and (16) which are quoted in the flat limits (cf. Sec. 5 below for details).

To establish the explicit (anti-)BRST invariance of the Lagrangian densities 
(7), (8), (15) and (16), we have to apply the (anti-)BRST transformations on every terms of the
above Lagrangian densities. This exercise is algebraically more involved as one has to collect the
terms containing $A_0, \, A_1, \, A_2, \, B_0, \, B_1$, separately and independently. In our Appendices B
and C, we have collected these terms which appear due to the applications of $s_B$ and
$\bar s_B$ on the Lagrangian densities ${\cal L}_{B}$ and ${\cal L}_{\bar B}$, respectively. The explicit form
of the BRST transformations on the BRST invariant Lagrangian density (7) (i.e. ${\cal L}_{B}$) is 
\begin{eqnarray}
s_B\, {\cal L}_{B} &=& \partial_a \, \Bigl[C^a \big({\cal L}_{0} + B_0\,A_0 + B_1\,A_1\big) + 
i\, \bar C_1 \, C^a (\partial_0 \, C^1 + \partial_1 \, C^0) \, A_0 \nonumber\\ &+& i\, \bar C_0 C^b\,\partial_b \, (C^a\,A_0)
+ i\,\bar C_0 \, C^a \,(\partial_0 \, C^1 + \partial_1 \, C^0) \, A_1 
+ i\, \bar C_1 C^b\,\partial_b \, (C^a\,A_1)\nonumber\\
&+& i\,\bar C_0 \, C^a \,(\partial_0 \, C^1 - \partial_1 \, C^0) \, A_2 +  
i\,\bar C_1 \, C^a \,(\partial_0 \, C^0 - \partial_1 \, C^1) \, A_2\Bigr]. 
\end{eqnarray}
In exactly similar fashion, the anti-BRST transformation acting on the anti-BRST
invariant Lagrangian density  ${\cal L}_{\bar B}$ produces the 
following explicit transformation: 
\begin{eqnarray}
\bar s_B\, {\cal L}_{\bar B} &=& \partial_a \, \Bigl[\bar C^a \big({\cal L}_{0} - \bar B_0\,A_0 - \bar B_1\,A_1\big) - 
i\, C_1 \, \bar C^a (\partial_0 \, \bar C^1 + \partial_1 \, \bar C^0) \, A_0 \nonumber\\
&-& i\, C_0 \,\bar C^b\,\partial_b \, (\bar C^a\,A_0)
- i\,C_0 \, \bar C^a \,(\partial_0 \, \bar C^1 + \partial_1 \, \bar C^0) \, A_1 
- i\,  C_1\,\bar C^b\,\partial_b \, (\bar C^a\,A_1)\nonumber\\
&-& i\, C_0 \, \bar C^a \,(\partial_0 \, \bar C^1 - \partial_1 \, \bar C^0) \, A_2 -  
i\, C_1 \, \bar C^a \,(\partial_0 \, \bar C^0 - \partial_1 \, \bar C^1) \, A_2\Bigr]. 
\end{eqnarray}
A close and careful look at (17) and (18) shows that we can obtain (18) from (17) provided we
make the replacements: $B_0 \to \bar B_0, \, B_1 \to \bar B_1, \, A_0 \to - A_0,\, A_1 \to - A_1,\,
 A_2  \to -  A_2, C_0 \leftrightarrow \bar C_0, \,  C_1 \leftrightarrow \bar C_1$.
Now it is obvious that, in the flat limit $A_0 = A_1 = 0, A_2 = 1$ of the full Lagrangian densities 
${\cal L}_{B}$ and ${\cal L}_{\bar B}$,  we obtain the following BRST and anti-BRST symmetry invariances
for the Lagrangian densities ${\cal L}_{B}^{(0)}$ and ${\cal L}_{\bar B}^{(0)}$, namely;
\begin{eqnarray}
s_B\, {\cal L}_{B}^{(0)} &=& \partial_a \, \Bigl[C^a \big({\cal L}_{0}\big) +
 i\,\bar C_0 \, C^a \,(\partial_0 \, C^1 - \partial_1 \, C^0)  +  
i\,\bar C_1 \, C^a \,(\partial_0 \, C^0 - \partial_1 \, C^1)\Bigr],
\end{eqnarray}
\begin{eqnarray}
\bar s_B\, {\cal L}_{\bar B}^{(0)} &=& \partial_a \, \Bigl[\bar C^a \big({\cal L}_{0}\big) 
- i\, C_0 \, \bar C^a \,(\partial_0 \, \bar C^1 - \partial_1 \, \bar C^0) -  
i\, C_1 \, \bar C^a \,(\partial_0 \, \bar C^0 - \partial_1 \, \bar C^1)\Bigr]. 
\end{eqnarray}
The total derivatives in (17), (18), (19) and (20) establish that the (anti-)BRST transformations
(12), (10), (9), (6) and (5) are the symmetries of the action integrals 
$S = \int d^2\xi \,{\cal L}_{B} $, $S = \int d^2\xi \,{\cal L}_{\bar B}$, 
$S = \int d^2\xi \,{\cal L}_{B}^{(0)} $ and $S = \int d^2\xi \,{\cal L}_{\bar B}^{(0)} $ 
provided we use the proper boundary conditions on the fields (and their derivatives) of the theory at 
$\sigma = 0 $ and $\sigma = \pi $ [6].

\section{Conserved Charges: Continuous Symmetries}

The BRST charge $Q_B$, that has been computed in [6], is in the flat limit ($A_0 = A_1 = 0, A_2 = 1$) where
the Lagrangian density ${\cal L}_B^{(0)}$ [cf. Eq. (8)] plays a pivotal role. First of all, we note that the Lagrangian
densities ${\cal L}_B^{(0)}$ and ${\cal L}_{\bar B}^{(0)}$ [cf. Eqs. (8) and (16)] respect the global ghost-scale 
symmetry transformations
\begin{eqnarray}
C_0 \to e^{\Omega} \, C_0, \qquad \bar C_0 \to e^{-\Omega} \, \bar C_0, \qquad
C_1 \to e^{\Omega} \, C_1   \qquad   \bar C_1 \to e^{-\Omega} \, \bar C_1,
\end{eqnarray}
where $\Omega$ is a global scale transformation parameter. For the sake of brevity, we set $\Omega = 1 $ so that the 
infinitesimal version ($s_g$) of the above global scale symmetry transformations  reduce to the following
transformations on the (anti-)ghost fields, namely;   
\begin{eqnarray}
s_g\, C_0 = C_0, \qquad s_g \,\bar C_0 = -\,\bar C_0, \qquad
s_g \,C_1 = C_1, \qquad      s_g \,\bar C_1 = - \, \bar C_1.
\end{eqnarray}
Here the subscript $g$ denotes the infinitesimal ghost scale transformations  (where $\Omega = 1$). The ghost charges, computed from 
${\cal L}_B^{(0)}$ and ${\cal L}_{\bar B}^{(0)}$,  are as follows   
\begin{eqnarray}
&&Q_g = \int_0^{\pi}  d\sigma\,J_g^{(0)} \equiv -\,i\,\int_0^{\pi} d\sigma\, (\bar C_0 \, C_1 - \bar C_1 \, C_0),\nonumber \\
&&\bar Q_g = \int_0^{\pi}  d\sigma\,\bar J_g^{(0)} \equiv -\,i\,\int_0^{\pi} d\sigma\, (\bar C_1 \, C_0 - \bar C_0 \, C_1),
\end{eqnarray}
where $J_g^{(0)}$ and $\bar J_g^{(0)}$ are the zeroth component of the Noether conserved currents
[corresponding to the infinitesimal ghost transformations (22)] that have been derived
from ${\cal L}_B^{(0)}$ and ${\cal L}_{\bar B}^{(0)}$, respectively. However, the above 
charges are {\it not} independent of each-other. Rather, they differ by a sign factor 
only (i.e. $Q_g = -\,\bar Q_g $). Using the following Euler-Lagrange equations of motion 
that emerge out from from  ${\cal L}_B^{(0)}$, namely;
\begin{eqnarray}
&&\square X^\mu = 0, \quad A_0 = A_1 = B_0 = B_1 = 0, \quad A_2 = 1, E = 0, \quad 
\partial_0 \bar C^0 + \partial_1 \bar C^1 = 0 \nonumber \\
&&\partial_0 C^0 - \partial_1 C^1 = 0,\quad \partial_0 C^1 - \partial_1 C^0 = 0, \quad
\partial_0 \bar C^1 + \partial_1 \bar C^0 = 0,
\end{eqnarray}
we observe that 
$\dot Q_g = i\,\int_0^{\pi} \frac {\partial} {\partial \sigma}\,\bigl[ \bar C_1\,C_0 - \bar C_0\,C_1 \bigr] = 0$
due to the boundary conditions. This shows that the ghost charge is conserved (i.e. $\dot Q_g = 0$).

We now concentrate on the derivation of the BRST charge $Q_B$ and anti-BRST charge $\bar Q_{\bar B}$
from the Lagrangian densities ${\cal L}_B^{(0)}$ and ${\cal L}_{\bar B}^{(0)}$, respectively. Taking into
account the basic concepts behind the Noether theorem, we note that  
$Q_B = \int_0^{\pi}  d\sigma\,J_B^{(0)},\, \bar Q_B = \int_0^{\pi}  d\sigma\,\bar J_B^{(0)}$  
where $J_B^{(0)}$ and $\bar J_B^{(0)}$ are the zeroth components of the Noether conserved currents
(corresponding to the BRST and anti-BRST  symmetry transformations) computed
from the Lagrangian densities ${\cal L}_B^{(0)}$ and ${\cal L}_{\bar B}^{(0)}$, respectively. 
The explicit expressions for these currents, derived from the above Lagrangian densities, are
\begin{eqnarray}
J_B^{(0)} &=& (s_B\, X^\mu) \, \frac{\partial {\cal L}_B^{(0)}}{\partial (\partial_0 X^\mu)}
+ (s_B\, C^0) \, \frac{\partial {\cal L}_B^{(0)}}{\partial (\partial_0 C^0)}
+ (s_B\, C^1) \, \frac{\partial {\cal L}_B^{(0)}}{\partial (\partial_0 C^1)} \nonumber\\
&+& (s_B\,\bar C_0) \, \frac{\partial {\cal L}_B^{(0)}}{\partial (\partial_0 \bar C_0)}
+ (s_B\, \bar C_1) \, \frac{\partial {\cal L}_B^{(0)}}{\partial (\partial_0 \bar C_1)} \, -\, X^0,
\end{eqnarray}
\begin{eqnarray}
\bar J_B^{(0)} &=& (\bar s_B\, X^\mu) \, \frac{\partial {\cal L}_{\bar B}^{(0)}}{\partial (\partial_0 X^\mu)}
+ (\bar s_B\, C^0) \, \frac{\partial {\cal L}_{\bar B}^{(0)}}{\partial (\partial_0 C^0)}
+ (\bar s_B\, C^1) \, \frac{\partial {\cal L}_{\bar B}^{(0)}}{\partial (\partial_0 C^1)} \nonumber\\
&+& (\bar s_B\,\bar C_0) \, \frac{\partial {\cal L}_{\bar B}^{(0)}}{\partial (\partial_0 \bar C_0)}
+ (\bar s_B\, \bar C_1) \, \frac{\partial {\cal L}_{\bar B}^{(0)}}{\partial (\partial_0 \bar C_1)} \, -\, Y^0,
\end{eqnarray}
where the explicit expressions for $X^0$ and $Y^0$ are:
\begin{eqnarray}
X^{(0)} &=& C^0\,{\cal L}_0 + i\,\bar C_0\, C^0 \,\big (\partial_0 C^1 - \partial_1 C^0 \big) 
+ i\,\bar C_1\, C^0 \,\big (\partial_0 C^0 - \partial_1 C^1 \big), \nonumber\\ 
Y^{(0)} &=& \bar C^0\,{\cal L}_0 - i\, C_0\, \bar C^0 \,\big (\partial_0 \bar C^1 - \partial_1 \bar C^0 \big) 
- i\, C_1\, \bar C^0 \,\big (\partial_0 \bar C^0 - \partial_1 \bar C^1 \big). 
\end{eqnarray}
The above expressions are derived from the equations (19) and (20) which are nothing  but the zeroth
components of the expressions that have been quoted in the square brackets. Finally, we obtain the following
expressions for the conserved BRST and anti-BRST charges  ($Q_B$ and $\bar Q_B$) from the
Lagrangian densities ${\cal L}_B^{(0)}$ and ${\cal L}_{\bar B}^{(0)}$, namely;
\begin{eqnarray}
Q_B &=& -\,\int_{0}^{\pi}\, d \sigma\, \Bigl [\frac{C^0}{2 \kappa} \, \big (\partial_0 X^\mu\, \partial_0 X_\mu  
+ \partial_1 X^\mu\, \partial_1 X_\mu \big )  + \frac{C^1}{2 \kappa} \, \big (\partial_0 X^\mu\, \partial_1 X_\mu  
+ \partial_1 X^\mu\, \partial_0 X_\mu \big ) \nonumber\\
&+& i\, \bar C_0 \, \big (C^a \, \partial_a C^1) +  i\, \bar C_1 \, \big (C^a \, \partial_a C^0) \Bigr ],
\end{eqnarray}
\begin{eqnarray}
\bar Q_B &=& -\,\int_{0}^{\pi}\, d \sigma\, \Bigl [\frac{\bar C^0}{2 \kappa} \, \big (\partial_0 X^\mu\, \partial_0 X_\mu  
+ \partial_1 X^\mu\, \partial_1 X_\mu \big )  + \frac{\bar C^1}{2 \kappa} \, \big (\partial_0 X^\mu\, \partial_1 X_\mu  
+ \partial_1 X^\mu\, \partial_0 X_\mu \big ) \nonumber\\
&+& i\, C_0  \big (\bar C^a  \partial_a \bar C^1) +  i\,  C_1  \big (\bar C^a  \partial_a \bar C^0) \Bigr ],
\end{eqnarray}
where we have used the Euler-Lagrange (EL) equations of motion (EoM) (24) derived from  
the Lagrangian density ${\cal L}_B^{(0)}$  and the following EL-EoM that emerge out from the 
Lagrangian density $ {\cal L}_{\bar B}^{(0)}$, namely;
\begin{eqnarray}
&&\Box X^\mu = 0, \qquad A_0 = A_1 = A_2 - 1 = \bar B_0 = \bar B_1 = E = 0, \qquad \partial_0 \bar C^0 
- \partial_1 \bar C^1 = 0,  \nonumber\\ 
&& \partial_0 \bar C^1 - \partial_1 \bar C^0 = 0, \qquad \qquad   \partial_0  C^1 + \partial_1 \bar C^0 = 0, \qquad
\qquad \partial_0 \bar C^0 + \partial_1 \bar C^1 = 0. 
\end{eqnarray}
In fact, a close and careful look at the EL-EoM (24) and (30) establishes the fact that $X^0 = Y^0 = 0$ on the
on-shell (because we substitute the EL-EoM into them).

The above charges $Q_B$ and $\bar Q_B$ are conserved. This can be checked by exploiting the
strength of the EL-EoM (24) and (30) while we take into account the direct ``time" derivative 
of the above charges, namely; 
\begin{eqnarray}
{\dot Q}_B &=& -\,\int_{0}^{\pi}\, d \sigma\, \frac{\partial}{\partial \sigma}\,\Bigl [\frac{C^0}{2 \kappa} \, 
\big (\partial_0 X^\mu\, \partial_0 X_\mu  
+ \partial_1 X^\mu\, \partial_1 X_\mu \big )  + \frac{C^1}{2 \kappa} \, \big (\partial_0 X^\mu\, \partial_1 X_\mu  
+ \partial_1 X^\mu\, \partial_0 X_\mu \big ) \nonumber\\
&+& i\, \bar C_0 \, \big (C^a \, \partial_a C^0) +  i\, \bar C_1 \, \big (C^a \, \partial_a C^1) \Bigr ],
\end{eqnarray}
\begin{eqnarray}
\dot {\bar Q}_B &=& -\,\int_{0}^{\pi}\, d \sigma\, \frac{\partial}{\partial \sigma}\,
 \Bigl [\frac{\bar C^0}{2 \kappa} \, \big (\partial_0 X^\mu\, \partial_0 X_\mu  
+ \partial_1 X^\mu\, \partial_1 X_\mu \big )  + \frac{\bar C^1}{2 \kappa} \, \big (\partial_0 X^\mu\, \partial_1 X_\mu  
+ \partial_1 X^\mu\, \partial_0 X_\mu \big ) \nonumber\\
&+& i\, C_0 \, \big (\bar C^a \, \partial_a \bar C^0) +  i\,  C_1 \, \big (\bar C^a \, \partial_a \bar C^1) \Bigr ].
\end{eqnarray}
The above expressions demonstrate that the BRST and anti-BRST charges are conserved when we use the boundary
conditions at $\sigma = 0$ and $\sigma = \pi$ on the appropriate fields and their derivatives 
(see, e.g. [6] for details). Thus, we have noted that there are {\it three} conserved charges
 (which correspond to {\it three} continuous symmetries that are present) in the theory. 
One can check, in a straightforward manner, that the ghost charge obeys the standard algebra with 
the BRST and anti-BRST charges. This can be checked in a simple manner by computing the left hand 
side of the following from (22), (23), (28) and (29), namely;
\begin{eqnarray}
s_g\,Q_g = -\,i\,[Q_g, \, Q_g] = 0, \;\; s_g\,Q_B = -\,i\,[Q_B, \, Q_g] = Q_B, 
\,\,\, s_g\,\bar Q_B = -\,i\,[Q_g, \, \bar Q_B] = - \, \bar Q_B,
\end{eqnarray}
which demonstrates that we have: $i\,[Q_g, \, Q_B] = +\,Q_B,$ and $i\,[Q_g, \, \bar Q_B] = -\,\bar Q_B,$. 
However, the proof of nilpotency of the 
BRST and anti-BRST charges requires very careful computations
at the {\it quantum} level where the normal mode expansions of the fields of our theory play very important
roles. In the paper by Kato and Ogawa [6], this exercise has been performed and it turns out that 
the nilpotency of the BRST charge is true {\it only} when D = 26 and $\alpha_0 = 1$. It is obvious that
we shall get the same result if we check the nilpotency of the anti-BRST charge at the quantum level
with the proper boundary conditions.

\section{Conclusions}

In our present investigation, we have been able to derive the {\it proper} anti-BRST symmetry 
transformations corresponding to the BRST transformations (that have been shown to be present
for the a model of 2D diffeomorphism invariant bosonic string theory [6]). The BRST and anti-BRST 
symmetry transformations 
are proved to be off-shell nilpotent of order two. However, these symmetries are found to be   
absolutely anticommuting {\it only} on a submanifold of the HIlbert space of quantum fields that is characterized 
by the 2D field equations (11). These {\it latter} equations are nothing but the CF-type restrictions 
which are the hallmark of the {\it quantum} diffeomorphism/gauge invariant theories when these theories are discussed
within the framework of BRST formalism. In fact, it is the existence of
the CF-type restrictions that primarily imply that the BRST and anti-BRST symmetries (and the
corresponding conserved charges) have their own identities. In the language of mathematics, 
they are linearly independent of each-other (on a submanifold of the quantum Hilbert space of fields)  that is defined by the
CF-type restrictions (11).

We have derived, in our present endeavor, the explicit and {\it separate} forms of BRST and anti-BRST invariant
Lagrangian densities and we have demonstrated {\it clearly} their transformation properties under the 
BRST and anti-BRST symmetry transformations. Using the Noether theorem, we have computed the
conserved BRST, anti-BRST and ghost charges of the theory in the {\it flat} limit. In fact, in the
latter limit, the BRST charge has {\it also} been derived by Kato and Ogawa [6]. We have shown that the 
standard algebra is obeyed between the ghost charge and BRST charge (as well as the ghost charge
and anti-BRST charge).  The nilpotency ($Q_B^2 = 0, {\bar Q}_B^2 = 0$) of the BRST ($Q_B$)  and 
anti-BRST (${\bar Q}_B$) charges has {\it not} been derived in our present investigation as this requires the
normal mode expansion of the fields and their substitution in the expressions for 
$Q_B$ and ${\bar Q}_B$. In fact, the requirement of the nilpotency of the BRST charge has led to
the derivation of D = 26 and $\alpha_0 = 1$ where D is the dimensionality of the target spacetime manifold 
and $\alpha_0$ is the intercept in the Regge trajectory that is generated 
due to the concept of strings (see, e.g. [6] for details).

We would like to comment a bit on the boundary conditions that are to be imposed on the fields
(and the derivatives on them) in our present theory when we demand the BRST as well as
anti-BRST invariance of the Lagrangian densities (7) and (15). For the BRST invariance of the theory,
the boundary conditions that have been obtained in the work by Kato and Ogawa [6] are: 
$\partial_1 \, X^\mu = 0, \, \bar C_0 = 0, C^1 = 0$ at $\sigma = 0$ and $\sigma = \pi$. 
The BRST invariance of the boundary condition $C^1 = 0$ (at $\sigma = 0$ and $\sigma = \pi$)
leads to the further boundary condition as: $\partial_0 \, C^1 = 0$ at $\sigma = 0$ and $\sigma = \pi$.
The anti-BRST invariance, in {\it exactly} similar manner, would lead to the boundary conditions
$\partial_1 \, X^\mu = 0, \,  C^0 = 0, \bar C_1 = 0$ at $\sigma = 0$ and $\sigma = \pi$. 
The anti-BRST invariance of the condition $\bar C_1 = 0$ at $\sigma = 0$ and $\sigma = \pi$ 
implies that  $\partial_0 \, \bar C_1 = 0$ (at $\sigma = 0$ and $\sigma = \pi$). Thus, the normal
mode expansions of the fields: $X^\mu (\tau, \sigma), \, C^0 (\tau, \sigma),  \, C^1 (\tau, \sigma),
 \, \bar C_0 (\tau, \sigma),  \, \bar C_1 (\tau, \sigma)$ can be found in the same manner
 as has been obtained in the work by Kato and Ogawa [6]. We have to be just careful that
for the anti-BRST invariance, the mode expansions in the ghost sector should be such that 
the expansions are exchanged, namely; $C_a \leftrightarrow \bar C_a$. The requirements of the
nilpotency of $Q_B$ and ${\bar Q}_B$ would obviously produce the results D = 26 and 
$\alpha_0 = 1$

We would like to mention that the BRST and anti-BRST invariant Lagrangian densities (7) and (15) 
have been derived in a straightforward manner by utilizing the gauge-fixing $A_0 = A_1 = 0$ and the 
(anti-)ghost fields [cf. Eqs. (4) and (14)]. However, if we compute the Lagrangian densities in the
Curci-Ferrari gauge [12,13] that would give due respect to the CF-type conditions that have been derived
in (11). We wish to devote time on the computation of the coupled Lagrangian densities (like 4D non-Abelian
gauge theory [10-13]) which produce the CF-type condition as the equations of motion. Furthermore,  the coupled
Lagrangian densities should respect {\it both} the BRST and anti-BRST symmetry transformations
on the submanifold of the quantum Hilbert space of fields  where teh specific quantum fields 
obey the CF-type restrictions (11). At present, we are working in this 
direction and our results would be reported elsewhere.

In a very recent work [7], the superfield approach to derive the proper (anti-)BRST symmetry
transformations for any general D-dimensional diffeomorphism invariant theory has been developed
(corresponding to its diffeomorphism symmetry invariance). It would be very nice future 
endeavor for us to apply the theoretical arsenals of this superfield formalism [7] to our present 
bosonic string model which is {\it also} a diffeomorphism invariant theory. In fact, we hope that
this superfield formalism would be able to shed more light on the geometrical origin and 
interpretation of the (anti-)BRST symmetries and the CF-type restrictions (11) which we have
obtained for our present theory. In our earlier works on the Abelian 2-form and 3-form gauge theories [14, 15], we have
established the geometrical origins and interpretations for the CF-type restrictions and their 
intimate connections with 
the geometrical objects called gerbes. 
It would be a challenging future endeavor for us to establish the {\it same} type of connections 
for our present 2D diffeomorphism invariant theory where the {\it non-trivial} CF-type restrictions exist. 
We are presently involved with this problem
and we plan to report about our progress in our future publication(s) [16].

It is gratifying to state that we have already exploited the beauty and strength of MBTSA in
the cases of 1D diffeomorphism (i.e. reparametrization) invarint interesting models of
non-relativistic free particle [17], scalar relativistic particle [18] and spinning (i.e. SUSY)
relativistic particle [19] and established the {\it universality} of the 1D CF-type restriction:
$B + \bar B + i\; (\bar C \dot C - \dot {\bar C} C) = 0$ in {\it all} the above
non-SUSY and SUSY systems of interest and obtained the proper (anti-)BRST symmetry
transformations corresponding to the {\it classical} 1D diffeomorphism (ie. reparameterization) 
symmetry transformation. Here the (anti-)ghost variables $(\bar C)C$ are the generalization 
of the infinitesimal reparametrization symmetry transformation parameter $\epsilon (\tau)$
in the transformation: $\tau \rightarrow \tau - \epsilon (\tau)$ where $\tau$ is an
evolution parameter (see, e.g. [17-19] for details).
In our present 2D diffoemorphism invariant bosonic string theory, we
have found the CF-type restrictions as: $B^a + \bar B^a + i \; (\bar C^b \;\partial_b \;C^a + C^b\; \partial_b \;\bar C^a) = 0$
(with $a, b, = 0, 1$) 
which are, once again, the limiting case of the MBTSA to D-dimensional diffeomorphism invariant
theory where it has been shown [7, 20] that the CF-type restrictions:  $B^\mu + \bar B^\mu 
+ i \;(\bar C^\rho\; \partial_\rho \;C^\mu + C^\rho \;\partial_\rho \;\bar C^\mu) = 0$ 
(with $\mu, \nu, \rho...= 0, 1, 2, ...D-1$) are {\it universal} for the 
SUSY and non-SUSY systems in any arbitrary dimension of spacetime where the (anti-)ghost fields
$(\bar C_\mu)C_\mu$ are the quantum generalizations of the infinitesimal D-dimensional
diffeomorphism transformation parameters $\varepsilon_\mu (x)$ in the transformations:
$x_\mu \rightarrow x_\mu - \varepsilon_\mu (x)$. In the above discussions, all the
$(\bar B)B$ fields, with appropriate index, are the Nakanishi-Lautrup type auxiliary fields. \\

\vskip 0.1cm

\begin{center}
{\bf Appendix A: On the Nilpotency Property $s_B^2 \, {\tilde g}^{ab} = 0$}\\
\end{center}
\vskip 0.2cm
\noindent
We briefly sketch here a few essentials steps that are needed in the proof of 
$s_B^2 \, {\tilde g}^{ab} = 0$. In this connection, we observe the following:
\begin{eqnarray}
s_B^2 \, {\tilde g}^{ab} = s_B\,\bigl[(\partial_m\,C^m) \, {\tilde g}^{ab}\bigr] + 
s_B\,\bigl[C^m \, \partial_m\, {\tilde g}^{ab}\bigr] - s_B\, \bigl[(\partial_m \, C^a)\, {\tilde g}^{mb}\bigr]
- s_B\, \bigl[(\partial_m \, C^b)\, {\tilde g}^{am}\bigr]. 
\end{eqnarray}
The {\it first} term, after the application of the BRST transformations, looks in its full glory as
\begin{eqnarray}
&&(\partial_m \, C^n) \, (\partial_n\,C^m)\,{\tilde g}^{ab} + C^m \, (\partial_m\,\partial_n \, C^n)\,{\tilde g}^{ab}
-  (\partial_m \, C^m) \, (\partial_n\,C^n)\,{\tilde g}^{ab} -
 (\partial_m \, C^m) \,C^n\,(\partial_n \, {\tilde g}^{ab})\nonumber\\ 
 && + (\partial_m \, C^m) \, (\partial_n\,C^a)\,{\tilde g}^{nb} +
 (\partial_m \, C^m) \, (\partial_n\,C^b)\,{\tilde g}^{an}.
\end{eqnarray}
In exactly similar fashion, the {\it second} term turns out to be 
\begin{eqnarray}
&&C^n\,(\partial_n \, C^m) \, (\partial_m\,{\tilde g}^{ab}) - C^m \, (\partial_m\,\partial_n \, C^n)\,{\tilde g}^{ab}
-  C^m\,(\partial_n \, C^n) \, (\partial_m\,{\tilde g}^{ab}) -
 C^m\,(\partial_m \, C^n) \,(\partial_n \, {\tilde g}^{ab})\nonumber\\ 
 && + C^n\, (\partial_n \, C^a) \, (\partial_m\,{\tilde g}^{mb}) +
C^n (\partial_n \, C^b) \, (\partial_m\,{\tilde g}^{an}) + C^m \, (\partial_m\,\partial_n \,C^a)\,{\tilde g}^{mb},
\end{eqnarray}
where we have taken into account the fact that $C^m\,C^n (\partial_m\,\partial_n\,{\tilde g}^{ab}) = 0$. 
The {\it third} term, after the application of the BRST transformations (5), (6) and (12), 
looks in the following exact mathematical form 
\begin{eqnarray}
&& - (\partial_m \, C^n) \, (\partial_n\,C^a)\,{\tilde g}^{mb} - C^n \, (\partial_m\,\partial_n \, C^a)\,{\tilde g}^{mb}
+ (\partial_m \, C^a) \, (\partial_n\,C^n)\,{\tilde g}^{mb} \nonumber\\ 
&& + (\partial_m \, C^a) \,C^n\,(\partial_n \, {\tilde g}^{mb}) - (\partial_m \, C^a) \, (\partial_n\,C^m)\,{\tilde g}^{nb} -
 (\partial_m \, C^a) \, (\partial_n\,C^b)\,{\tilde g}^{mn}.
\end{eqnarray}
Finally, the {\it fourth} term can be explicitly expressed,  after the application of BRST 
transformations (5), (6) and (12), as 
\begin{eqnarray}
&& - (\partial_m \, C^n) \, (\partial_n\,C^b)\,{\tilde g}^{am} - C^n \, (\partial_m\,\partial_n \, C^b)\,{\tilde g}^{am}
+ (\partial_m \, C^b) \, (\partial_n\,C^n)\,{\tilde g}^{am} +
 (\partial_m \, C^b) \,C^n\,(\partial_n \, {\tilde g}^{am})\nonumber\\ 
 && - (\partial_m \, C^b) \, (\partial_n\,C^m)\,{\tilde g}^{an} -
 (\partial_m \, C^b) \, (\partial_n\,C^a)\,{\tilde g}^{mn}.
\end{eqnarray}
It is evident that the following terms from (35), (36), (37) and (38), namely;
\begin{eqnarray}
&& C^n \, (\partial_m\,\partial_n \, C^m)\,{\tilde g}^{ab} - C^m \, (\partial_m\,\partial_n \, C^n)\,{\tilde g}^{ab} 
+ C^m \, (\partial_m\,\partial_n \, C^a)\,{\tilde g}^{mb}\nonumber\\
&&  C^m \, (\partial_m\,\partial_n \, C^b)\,{\tilde g}^{an} - C^n \, (\partial_m\,\partial_n \, C^a)\,{\tilde g}^{mb} 
- C^m \, (\partial_m\,\partial_n \, C^b)\,{\tilde g}^{an}, 
\end{eqnarray}
cancel out with one-another. Furthermore, the following terms from (37) and (38), 
\begin{eqnarray}
- (\partial_m \, C^a) \, (\partial_n\,C^b)\,{\tilde g}^{mn} - (\partial_m\, C^b )\,(\partial_n\,C^a)\,{\tilde g}^{mn},
\end{eqnarray}
cancel out with each-other because of the antisymmetric nature ($C^a\,C^b + C^b\,C^a = 0$) of
the ghost fields ($C^a$) and the symmetric nature ($ {\tilde g}^{mn} = {\tilde g}^{nm}$) 
of the metric tensor ${\tilde g}^{mn}$. Rest of the terms also cancel out by taking the 
help of the exchange of dummy indices $m \leftrightarrow n$ and the anticommuting nature 
of the ghost fields. Finally, we find that the following terms, from the sum of 
(35), (36), (37) and (38), remain left-out at the end, namely;
\begin{eqnarray}
\Big[(\partial_n \, C^m) \, (\partial_m\,C^n) - (\partial_m\, C^m )\,(\partial_n\,C^n)\Big]\,{\tilde g}^{ab}.
\end{eqnarray}
The terms in the square bracket turn out to be {\it individually} equal to zero when we take the sum over $m, n = 0, 1$. 
This establishes the  nilpotency ($s_B^2 = 0$) of $s_B$ when it acts on ${\tilde g}^{ab}$.

\vskip 1.5cm

\begin{center}
{\bf Appendix B: On the BRST Symmetry Invariance of ${\cal L}_B$}\\
\end{center}
\vskip 0.5cm
\noindent
We collect here all the terms that are generated due to the application of BRST symmetry
transformations ($s_B$) on ${\cal L}_B$ [cf. Eq. (7)]. It is straightforward to note that
$s_B\, {\cal L}_0 = \partial_a\,(C^a\,{\cal L}_0)$. We assemble, first of all, the terms that contain
$ B_0 $ and $ B_1 $ fields due to the application of $s_B $ on {\it all} the terms that are present in 
${\cal L}_B$. These terms with $B_0$ field  are 
\begin{eqnarray}
&& C^a\,(\partial_a\,B_0)\,A_0 + B_0\,C^a\,(\partial_a\,A_0) - B_0\,(\partial_0 C^1 + \partial_1 C^0)\,A_1\nonumber\\
&& - B_0\, (\partial_0 C^1 - \partial_1 C^0)\,A_2 
 - B_0\,(\partial_1 C^0)\,A_2  + B_0\,(\partial_0 C^1)\,A_2  \nonumber\\ && + B_0\,(\partial_1 C^0)\,A_1 
+ B_0\,(\partial_0 C^1)\,A_1 + B_0\,(\partial_a C^a)\,A_0.
\end{eqnarray}
Similarly, the terms containing $B_1$ fields are as follows:
\begin{eqnarray}
&&  B_1\,C^a\,(\partial_a A_1) - B_1\,(\partial_a C^a)\,A_2 + B_1\,(\partial_1 C^1)\,A_2 
- B_1\,(\partial_1 C^0)\,A_0 \nonumber\\ 
&& + B_1\,(\partial_a C^a)\,A_1 - B_1\,(\partial_0 C^1)\,A_0 
 - B_1\,(\partial_1 C^1)\,A_2 + B_1\,(\partial_0 C^0)\,A_2  \nonumber\\
&& + B_1\,(\partial_0 C^1)\,A_0 + B_1\,(\partial_1 C^0)\,A_0 
+ C^a\,(\partial_a B_1)\,A_1.
\end{eqnarray}
It is clear that if we sum these terms [i.e. (42) and (43)] carefully with 
$s_B\, {\cal L}_0 = \partial_a\,(C^a\,{\cal L}_0)$, 
they lead to the sum of the following total derivative:
\begin{eqnarray}
\partial_a \, \Big[C^a \, \big({\cal L}_0 + B_0\,A_0 + B_1 \, A_1\big)\Big].
\end{eqnarray}
Thus far, we have obtained the total derivative from the original Lagrangian density (1) and 
terms that contain {\it necessarily} the Nakanishi-Lautrup auxiliary fields $B_0$ and $B_1$. It is 
straightforward to note that the original Lagrangian density (1) respects the {\it classical}
2D diffeomorphism symmetry transformations that have been pointed out in our Sec. 2. Thus, 
it is crystal clear that it will respect the BRST symmetry transformation where the
{\it classical} diffeomorphism transformation parameter $\varepsilon^a$ is replaced by
the {\it quantum} ghost field $C^a$. This is why we have $s_B {\cal L}_0 = \partial_a [C^a  {\cal L}_0]$
which is explicitly present in Eq. (44).

We now collect the terms that incorporate $A_2$ after the application of $s_B$ on 
${\cal L}_B$ [cf. Eq. (7)]. These useful terms are as follows:
\begin{eqnarray}
&&i\, \bar C_1\,(\partial_1\,C^1)\, C^ a\, (\partial_a\,A_2) \, - i\, C^ a\,(\partial_a\,\bar C_0)\,
(\partial_0\,C^1 - \partial_1\,C^0)\,A_2  - i\, \bar C_1 \,(\partial_0\,C^0)\, C^a \partial_a\,A_2 \nonumber\\
&&+ i\, \bar C_0\,(\partial_1\,C^0)\, C^ a\, (\partial_a\,A_2) \, - i\, \bar C_o (\partial_0 \, C^1)\, 
C^a \, (\partial_a \, A_2) + i\, \bar C_1 (\partial_a\,C^a) \, (\partial_0 \, C^0 - \partial_1\, C^1)\,A_2 \nonumber\\
&& - i\, C^a \,(\partial_a \, \bar C_1) \, (\partial_0 \, C^0 - \partial_1 \, C^1)\, A_2 
+ i\,\bar C_0 \, (\partial_a\,C^a)\, (\partial_0 \, C^1 - \partial_1\,C^0)\, A_2
 - i\,\bar C_1 \, C^a\,(\partial_a \,\partial_1\,C^1)\, A_2  \nonumber\\
&& + i\,\bar C_1 \, (\partial_0\,C^1)\, (\partial_0 \, C^1 - \partial_1\,C^0)\, A_2 + 
i\,\bar C_1 \, (\partial_1\,C^0)\, (\partial_0 \, C^1 - \partial_1\,C^0)\, A_2 
- i\,\bar C_1 \, (\partial_1\,C^a)\, (\partial_a \, C^1)\, A_2\nonumber\\ 
&& - i\,\bar C_1 \, (\partial_0\,C^a)\, (\partial_a \, C^0)\, A_2 
+ i\,\bar C_1 \, C^a\,(\partial_a \,\partial_0\,C^0)\, A_2 
+ i\,\bar C_0 \, (\partial_0\,C^1)\, (\partial_0 \, C^0 - \partial_1\, C^1)\, A_2 \nonumber\\
&& - i\,\bar C_0 \, (\partial_1\,C^a)\, (\partial_a \, C^0)\, A_2 
- i\,\bar C_0 \, C^a\,(\partial_a \,\partial_1\,C^0)\, A_2
+ i\,\bar C_0 \, (\partial_0\,C^a)\, (\partial_a \, C^1)\, A_2 \nonumber\\ 
&& + i\,\bar C_0 \, C^a\,(\partial_a \,\partial_0\,C^1)\, A_2
+ i\,\bar C_0 \, (\partial_1\,C^0)\, (\partial_0 \, C^0 - \partial_1\, C^1)\, A_2.
\end{eqnarray}
It is very interesting to note that all these terms, after many surprising cancellations,
sum-up to yield a total derivative as:
\begin{eqnarray}
\partial_a \, \Big[i\, \bar C_0 \, C^a \, \big(\partial_0\,C^1 - \partial_1\,C^0 \big)\,A_2
+ i\, \bar C_1 \, C^a \, \big(\partial_0\,C^0 - \partial_1\,C^1 \big)\,A_2\Big].
\end{eqnarray}
We now concentrate on all the terms that contain $A_0$ which emerge out from the application 
of $s_B$ on the relevant terms of the Lagrangian density ${\cal L}_B$. These are as follows:
\begin{eqnarray}
&& i\, \bar C_1 \, (\partial_0\,C^0) \, (\partial_1\,C^0 - \partial_0\,C^1) \, A_0
- i\, \bar C_1 \, (\partial_1\,C^1) \, (\partial_1\,C^0 - \partial_0\,C^1) \, A_0\nonumber\\ 
&& - i\, \bar C_0 \, (\partial_1\,C^0) \, (\partial_1\,C^0 - \partial_0\,C^1) \, A_0
 + i\, \bar C_0 \, (\partial_0\,C^1) \, (\partial_1\,C^0 - \partial_0\,C^1) \, A_0\nonumber\\ 
&& + i\, \bar C_0 \, (\partial_1\,C^0) \, (\partial_1\,C^0 + \partial_0\,C^1)\, A_0
+ i\,\bar C_0 \, (\partial_0\,C^1) \, (\partial_1\,C^0 + \partial_0\,C^1)\, A_0
\nonumber\\
&& + i\, \bar C_1 \, C^a \, (\partial_1\,\partial_a\,C^0) \, A_0
- i\, \bar C_1 \, (\partial_1\,C^0) \, C^a\, \partial_a\,A_0
+ i\, \bar C_1 \, (\partial_0 \,C^a) \, (\partial_a\,C^1) \, A_0\nonumber\\
&& + i\, \bar C_1 \, C^a \, (\partial_0\,\partial_a\,C^1) \, A_0
- i\, \bar C_1 \, (\partial_0\,C^1) \, C^a\, \partial_a\,A_0
+ i\, \bar C_0 \,(\partial_a \,C^b)\, (\partial_b\,C^a) \, A_0\nonumber\\
&&  + i\, \bar C_0 \, C^b \, (\partial_b\,\partial_a\,C^a) \, A_0
- i\, \bar C_0 \,(\partial_a \,C^a)\, C^b \,\partial_b \, A_0
+ i\, \bar C_b \,(\partial_b \,C^a)\, (\partial_a\,\bar C_0) \, A_0\nonumber\\
&& + i\, C^a \,(\partial_a\,\bar C_0) \,C^b \,\partial_b \, A_0
+ i\, \bar C_1 \, (\partial_1\,C^a) \, (\partial_a\,C^0) \, A_0.
\end{eqnarray}
It is amazing to find out that the sum of the above terms, after some miraculous
cancellations, yields a total derivative as: 
\begin{eqnarray}
\partial_a\,[i\,\bar C_1 \,C^a \,(\partial_0\,C^1 + \partial_1\,C^0)\,A_0 
+ i\,\bar C_0 \,C^b \,\partial_b (C^a \, A_0)].
\end{eqnarray}
Now, at the fag end, we have only one option left-out. As a consequence, ultimately,
we focus on the terms that necessarily incorporate $A_1$ field after the application
of the BRST transformation $s_B$ on  the Lagrangian density (7). These terms are: 
\begin{eqnarray}
&& i\, \bar C_0 \, (\partial_a\,C^a) \, (\partial_0\,C^1 + \partial_1\,C^0) \, A_1
 + i\, \bar C_1 \, C^b\, \partial_b \, \partial_a\,C^a \, A_1
 - i\, \bar C_1 \, (\partial_a\,C^a) \, C^b\, \partial_b \, A_1\nonumber\\
&& + i\, C^b\, \partial_b \,C^a \,(\partial_a\,\bar C_1) \, A_1
+ i\, C^b\, (\partial_b \,\bar C_1) \,C^a\,\partial_a\, A_1
- i\, \bar C_0 \, (\partial_0\,C^1) \, C^a\,\partial_a\, A_1\nonumber\\
&&+ i\, \bar C_1 \, (\partial_a\,C^b) \, C^a\,\partial_a\,A_1
- i\, \bar C_0 \, (\partial_1\,C^0) \,  C^a\,\partial_a\,A_1
 + i\,\bar C_0 \, C^a\,(\partial_0 \,\partial_a\,C^1)\, A_1 \nonumber\\
&& - i\, \bar C_1 \, (\partial_1\,C^1) \, (\partial_0\,C^0 - \partial_1\,C^1) \, A_1
+ i\, \bar C_1 \, (\partial_0\,C^0) \, (\partial_0\,C^0 - \partial_1\,C^1) \, A_1\nonumber\\ 
&& - i\, \bar C_0 \, (\partial_1\,C^0) \, (\partial_0\,C^0 - \partial_1\,C^1) \, A_1
 + i\, \bar C_0 \, (\partial_0\,C^1) \, (\partial_0\,C^0 - \partial_1\,C^1) \, A_1\nonumber\\ 
&& - i\, \bar C_0 \, (\partial_1\,C^a) \, (\partial_a\,C^0)\, A_1
+ i\,\bar C_0 \, C^a\,(\partial_1 \,\partial_a\,C^0)\, A_1
+ i\, \bar C_0 \, (\partial_0\,C^a) \, (\partial_a\,C^1)\, A_1 \nonumber\\ 
&& + i\, \bar C_1 \, (\partial_0\,C^1) \, (\partial_0\,C^1 + \partial_1\,C^0) \, A_1 
+ i\, \bar C_1 \, (\partial_1\,C^0) \, (\partial_0\,C^1 + \partial_1\,C^0) \, A_1 \nonumber\\ 
&& - i\,  C^a \, (\partial_a\,\bar C_0) \, (\partial_0\,C^1 + \partial_1\,C^0) \, A_1.
\end{eqnarray}
The above terms add-up to yield a total derivative term as:
\begin{eqnarray}
\partial_a\,[i\,\bar C_0 \,C^a \,(\partial_0\,C^1 + \partial_1\,C^0)\,A_1 
+ i\,\bar C_1 \,C^b \,\partial_b (C^a \, A_1)]. 
\end{eqnarray}
It is interesting to point out that the terms with $A_0$ and that of $A_1$ sum-up to yield exactly similar 
types of result in the total derivative where $A_0 \leftrightarrow A_1, \, 
\bar C_0 \leftrightarrow \bar C_1$. It is clear that the application of $s_B$
on ${\cal L}_B$ produces the total derivative term which is the sum of (44), (46), (48)
and (50). Thus, the BRST transformations $s_B$ is a {\it symmetry} of the action.

\vskip 1cm

\begin{center}
{\bf Appendix C: On the Anti-BRST Symmetry Invariance of ${\cal L}_{\bar B}$}\\
\end{center}
\vskip 0.5cm
\noindent
We collect here the terms that are generated after the application of the anti-BRST symmetry
transformations $\bar s_B$ on  ${\cal L}_{\bar B}$ [cf. Eq. (15)]. It can be readily checked 
that $\bar s_B\, {\cal L}_{0} = \partial_a\,(\bar C^a\,{\cal L}_0)$. In addition to it, we have
the following terms that contain the auxiliary field $\bar B_0$ after the application
of $\bar s_B$ on $ {\cal L}_{\bar B}$, namely; 
\begin{eqnarray}
&& - \bar C^a\,(\partial_a\,\bar B_0)\,A_0 - \bar B_0\,\bar C^a\,(\partial_a\,A_0) 
+ \bar B_0\,(\partial_0 \, \bar C^1 + \partial_1 C^0)\,A_1\nonumber\\
&& + \bar B_0\, (\partial_0 \, \bar C^1 - \partial_1 \, \bar C^0)\,A_2 
 + \bar B_0\,(\partial_1 \, \bar C^0)\,A_2  - \bar B_0\,(\partial_0 \, \bar C^1)\,A_2  \nonumber\\ 
&& - \bar B_0\,(\partial_1 \, \bar C^0)\,A_1 
- \bar B_0\,(\partial_0 \, \bar C^1)\,A_1 - \bar B_0\,(\partial_a \, \bar C^a)\,A_0,
\end{eqnarray}
which add-up to yield $ \partial_a \, [ -\,\bar C^a \, \bar B_0\,A_0)]$. Similarly, the 
following terms containing $\bar B_1$ fields (that are generated after the application
of $\bar s_B$ on  ${\cal L}_{\bar B}$), namely; 
\begin{eqnarray}
&& - \bar B_1\,\bar C^a\,(\partial_a A_1) + \bar B_1\,(\partial_0 \, \bar C^0)\,A_2 -
\bar B_1\,(\partial_1 \, \bar C^1)\,A_2 + \bar B_1\,(\partial_1 \, \bar C^0)\,A_0 \nonumber\\ 
&& - \bar B_1\,(\partial_a \, \bar C^a)\,A_1 + \bar B_1\,(\partial_0\, \bar C^1)\,A_0 
 + \bar B_1\,(\partial_1 \, \bar C^1)\,A_2 - \bar B_1\,(\partial_0 \, \bar C^0)\,A_2  \nonumber\\
&& - \bar B_1\,(\partial_0 \, \bar C^1)\,A_0 - \bar B_1\,(\partial_1 \,  \bar C^0)\,A_0 
- \bar C^a\,(\partial_a \, \bar B_1)\,A_1,
\end{eqnarray}
sum-up to produce $ \partial_a \, [ -\,\bar C^a \, \bar B_1\,A_0]$. Thus, it is clear that
we have so far the following total derivatives:
$ \partial_a \,[\bar C^a \, ({\cal L}_0 - \bar B_0\,A_0 - \bar B_1 \, A_1)]$.
Now we focus on the collection of $A_0$ terms that are generated after the application of 
anti-BRST transformations $\bar s_B$ on  ${\cal L}_{\bar B}$. These are 
\begin{eqnarray}
&&+ \,i\,\partial_a\, (C_0\,\bar C^a) \, (\bar C^b \, \partial_b\,A_0)
+ \,i\,C_1\, (\partial_1\, \bar C^0 + \partial_0\, \bar C^1) \,(\bar C^b \, \partial_b\,A_0)\nonumber\\
&&-\,\partial_a\, \big[i\,C_0 \, (\bar C^b \, \partial_b\, \bar C^a)\big]\,A_0 
-\,i\,C_1\, \partial_0\, (\bar C^b \, \partial_b\, \bar C^1)\,A_0 
-\,i\,C_1\, \partial_1\, (\bar C^b \, \partial_b\, \bar C^0)\,A_0 \nonumber\\
&& - \,i\,\partial_a\, (C_1\,\bar C^a) \, (\partial_1\, \bar C^0 + \partial_0\, \bar C^1)\,A_0 
 -\,i\,C_1\,(\partial_0\, \bar C^0 - \partial_1\, \bar C^1) \,(\partial_1\, \bar C^0 - \partial_0\, \bar C^1)\,A_0.
\end{eqnarray}
It will be noted that we have collected here the $A_0$ terms which look completely different 
from the corresponding terms in the BRST symmetry invariance [cf. Eq. (47)]. This is due to the 
fact we have not written each term separately and independently. However, these terms are actually 
similar to (47). The above terms add-up to produce the following total derivative terms, namely, 
\begin{eqnarray}
\partial_a \, \big[ -i\, C_1\,\bar C^a \, (\partial_0\, \bar C^1 + \partial_1\,\bar C^0)\,A_0
- i\,C_0\,\bar C^b\,\partial_b\, (\bar C^a\,A_0)\big].
\end{eqnarray}
We now concentrate on all the terms that are generated after the application of $\bar s_B$ on 
${\cal L}_{\bar B}$ and contain {\it necessarily} $A_1$ field. These are as follows
\begin{eqnarray}
&&+ \,i\,\partial_a\, (C_1\,\bar C^a) \, (\bar C^b \, \partial_b\,A_1)
+ \,i\,C_0\, (\partial_1\, \bar C^0 + \partial_0\, \bar C^1) \,(\bar C^b \, \partial_b\,A_1)\nonumber\\
&&-\,\partial_a\, \big[i\,C_1 \, (\bar C^b \, \partial_b\, \bar C^a)\big]\,A_1 
-\,i\,C_0\, \partial_0\, (\bar C^b \, \partial_b\, \bar C^1)\,A_1 
-\,i\,C_0\, \partial_1\, (\bar C^b \, \partial_b\, \bar C^0)\,A_1 \nonumber\\
&& - \,i\,\partial_a\, (C_0\,\bar C^a) \, (\partial_1\, \bar C^0 + \partial_0\, \bar C^1)\,A_1 
 -\,i\,C_1\,(\partial_0\, \bar C^0 - \partial_1\, \bar C^1) \,(\partial_1\, \bar C^0 - \partial_0\, \bar C^1)\,A_1.
\end{eqnarray}
The above terms add-up to produce the following total derivative
\begin{eqnarray}
\partial_a \, \big[ -i\, C_0\,\bar C^a \, (\partial_0\, \bar C^1 + \partial_1\,\bar C^0)\,A_1
- i\,C_1\,\bar C^b\,\partial_b\, (\bar C^a\,A_1)\big].
\end{eqnarray}
Finally, we have the following set of terms that contain {\it necessarily} $A_2$ field after the application of 
$\bar s_B$ on ${\cal L}_{\bar B}$, namely;
\begin{eqnarray}
&&-\,\partial_a\,[i\,C_0 \, \bar C^a]\,(\partial_0\, \bar C^1 - \partial_1\, \bar C^0) A_2 
-\,i\,C_0\, (\partial_0\, \bar C^1 + \partial_1\, \bar C^0)\, (\partial_0\, \bar C^0 - \partial_1\, \bar C^1)\,A_2\nonumber\\
&& +\,i\,C_0\,(\partial_0\, \bar C^1 - \partial_1\, \bar C^0)\,(\bar C^b \, \partial_b\,A_2)
- \,i\,\partial_a\, (C_1\,\bar C^a) \, (\partial_0\, \bar C^0  - \partial_1\, \bar C^1)\,A_2\nonumber\\
&& + \,i\,C_1\, (\partial_0\, \bar C^0 - \partial_1\, \bar C^1) \,(\bar C^b \, \partial_b\,A_2)
- \,i\,C_0\, \partial_0\, (\bar C^b \, \partial_b\, \bar C^1)\,A_2
 -\,i\,C_1\,\partial_0\,(\bar C^b \, \partial_b\,\bar C^0)\,A_2 \nonumber\\ 
 && + i\,C_a\, \partial_1\, (\bar C^b \,\partial_b\, \bar C^a)\,A_2
 + 2\,i\,C_1 \, \partial_0\, \bar C^1) \, (\partial_1\, \bar C^0) \, A_2.
\end{eqnarray}
The above terms produce, after their addition, the following total derivative:
\begin{eqnarray}
\partial_a \, [ -i\, C_0\,\bar C^a \, (\partial_0\, \bar C^1 - \partial_1\,\bar C^0)\,A_2
- i\,C_1\,\bar C^a\,(\partial_0\, \bar C^0 - \partial_1\,\bar C^1)\,A_2].
\end{eqnarray}
The sum of all the total derivatives, present in this Appendix, sum-up to produce
the total derivative that has been quoted in the main body of our text [cf. Eq. (18)].

\vskip 1cm

\begin{center}
{\bf Appendix D: On the Superfield Approach to the Derivation of the CF-Type Restrictions and Absolute Anticommutativity}\\
\end{center}
\vskip 0.5cm

\noindent
In this Appendix, we exploit the theoretical tricks and strength of the {\it modified} version of Bonora-Tonin superfield approach (MBTSA)
to BRST formalism [7] to concisely derive the 2D version of the {\it universal} CF-type restrictions: $B^a + \bar B^a + i\,(C^m\,\partial_m\,\bar C^a + \bar C^m\,\partial_m\,
C^a) = 0$. In this context, first of all, we focus on the {\it general} form of the 2D diffeomorphism transformations: $\xi^a \rightarrow \xi^{'a} = 
h^a(\xi)$ where $h^a(\xi)$ is a physically {\it well-defined} function of $\xi^a$ such that it is {\it finite} at the origin 
and vanishes off at $\tau \rightarrow \pm\,\infty$. We take now the generalizations of function $h^a(\xi)$ onto the $(2, 2)$-dimensional
supermanifold as follows
\[
h^a(\xi) \quad \rightarrow \quad \tilde h^a(\xi, \theta, \bar\theta) = \xi^a - \theta\,\bar C^a (\xi) - \theta\,C^a (\xi) + \theta\,\bar\theta\,k^a(\xi), \eqno(D.1)
\]  
where the (anti-)ghost fields $(\bar C^a)C^a$ are the {\it ones} which have {\it appeared} in the (anti-)BRST symmetry transformations (6) and (9).
We note, at this stage, that the infinitesimal version of the {\it classical} diffeomorphism transformations: $\xi^a \rightarrow \xi^{'a} = h^a(\xi) \equiv 
\xi^a - \varepsilon^a(\xi)$ denotes that the {\it classical} diffeomorphism transformations: $\delta\,\xi^a = \xi^{'a} - \xi^a = \varepsilon^a(\xi)$ can be 
{\it elevated} to their counterpart {\it quantum} (anti-)BRST
symmetry transformations as: $s_b\,\xi^a = -\,C^a, \, s_{ab}\,\xi^a = -\,\bar C^a$. This is the reason behind the choice in (D.1) where the 
coefficients of $\theta$ and $\bar\theta$ are $\bar C^a$ and $C^a$, respectively. This has been done due to the standard BRST prescription where
the infinitesimal {\it bosonic} parameters $\varepsilon^a(\xi)$ are replaced by the {\it fermionic} [i.e. $(C^a)^2 = (\bar C^a)^2 = 0,\, 
C^a\,\bar C^b + \bar C^b\,C^A = 0,$ etc] (anti-)ghost fields $(\bar C^a)C^a$ within the framework of
BRST formalism. The secondary fields $k^a(\xi)$ [i.e. the coefficient of $\theta\,\bar\theta$ in (D.1)] have to be determined from 
the consistency conditions and basic concepts behind  the BRST formalism (which we elaborate below in a very concise manner).

It is the basic tenet of MBTSA that the target space coordinate fields $X^\mu(\xi)$ have to be generalized onto their counterpart superfields on
the $(2, 2)$-dimensional supermanifold as:
\[
X^\mu(\xi) \quad \rightarrow \quad \tilde X^\mu\big[\tilde h(\xi, \theta, \bar\theta), \theta, \bar\theta\big] = 
{\cal X}^\mu[\tilde h(\xi, \theta, \bar\theta)] + \theta\,\bar Q^\mu[\tilde h(\xi, \theta, \bar\theta)] 
\]
\[
+ \bar\theta\,Q^\mu[\tilde h(\xi, \theta, \bar\theta)]
+ \theta\,\bar\theta\,T^\mu[\tilde h(\xi, \theta, \bar\theta)].     \eqno(D.2)
\]    
It should be noted, at this juncture, that all the secondary superfields on the r.h.s. are {\it still} function of transformation
$\tilde h^a(\xi, \theta, \bar\theta)$ that has been pointed out in (D.1). We take {\it another} Taylor expansion for all the secondary
superfields on the r.h.s. as:
\[
\theta\,\bar\theta\,T^\mu[\xi^a - \theta\,\bar C^a - \bar\theta\,C^a + \theta\,\bar\theta\,k^a] = \theta\,\bar\theta\,T^\mu(\xi),
\]  
\[
\bar\theta\,Q^\mu[\xi^a - \theta\,\bar C^a - \bar\theta\,C^a + \theta\,\bar\theta\,k^a] = \bar\theta\,Q^\mu(\xi) 
+ \theta\,\bar\theta\,\bar C^a\,
\partial_a\,Q^\mu(\xi),
\]
\[
\theta\,\bar Q^\mu[\xi^a - \theta\,\bar C^a - \bar\theta\,C^a + \theta\,\bar\theta\,k^a] 
= \theta\,\bar Q^\mu(\xi) - \theta\,\bar\theta\,C^a\,
\partial_a\,\bar Q^\mu(\xi),
\]
\[
{\cal X}^\mu[\xi^a - \theta\,\bar C^a - \bar\theta\,C^a + \theta\,\bar\theta\,k^a] = X^\mu(\xi) - \theta\,\bar C^a\,\partial_a\,X^\mu 
- \bar\theta\,C^a\,\partial_a\,X^\mu 
\]
\[
+ \theta\,\bar\theta\,\big[k^a\,\partial_a\,X^\mu - \bar C^a\,C^b\,\partial_a\,\partial_b\,X^\mu\big]. \eqno(D.3)
\]
It is pertinent and important to point out that we have to make {\it two-step} generalizations of $X^\mu(\xi)$ to the $(2, 2)$-dimensional
supermanifold in order to incorporate the diffeomorphism symmetry transformation $h^a(\xi) \rightarrow \tilde h^a(\xi, \theta, \bar\theta)$.
In other words, we have the following 
\[
X^\mu(\xi) \quad \rightarrow \quad {\cal X}^\mu(\xi, \theta, \bar\theta) \quad \rightarrow \quad {\tilde X}^\mu[\tilde h(\xi, \theta, \bar\theta), 
\theta, \bar\theta], \eqno(D.4)
\]
where, in the {\it first-step}, the target space coordinate field $[X^\mu(\xi)]$ has been generalized to ${\cal X}^\mu(\xi, \theta, \bar\theta)$
when there is {\it no} diffeomorphism transformation on $\xi^a$. In the next step, we incorporate $\xi^a \rightarrow \tilde h^a(\xi, \theta, \bar\theta)$, and
then, we make the super expansion as given in (D.2). Collecting the coefficients of $\theta, \,\bar\theta$ and $\theta\,\bar\theta$
from the r.h.s. [cf. Eq. (D.3)], we obtain
\[
\tilde X^\mu[\tilde h(\xi, \theta, \bar\theta), \theta, \bar\theta] = X^\mu(\xi) + \theta \,[\bar Q^\mu - \bar C^a\,\partial_a\,X^\mu] 
+  \bar\theta \,[Q^\mu - C^a\,\partial_a\,X^\mu]
\]
\[
+ \theta \, \bar\theta  [T^\mu + \bar C^a \, \partial_a\,Q^\mu - C^a\,\partial_a\,\bar Q^\mu + k^a\,\partial_a\,X^\mu - 
\bar C^a\,C^b\,\partial_a\,\partial_b\,X^\mu]. \eqno(D.5)
\]
At this stage, our key objective is to determine the values of $Q^\mu,\,\bar Q^\mu$ and $T^\mu$ so that we can obtain the {\it quantum}
(anti-)BRST symmetries (corresponding to the 2D {\it classical} diffeomorphism symmetry transformations) for the target space coordinates
$X^\mu(\xi)$ {\it and} the CF-type restrictions: $B^a + \bar B^a + i\, \big(C^m \, \partial_m \, \bar C^a + \bar C^m \, \partial_m \, C^a \big) = 0$.

To accomplish the above goal, we apply the horizontality condition (HC)
\[
X^\mu\,[\tilde h^a(\xi,\theta, \bar\theta)] = X^\mu (\xi), \eqno(D.6)
\]
which amounts to setting the coefficients of $\theta$, $\bar\theta$ and $\theta \,\bar\theta$ equal to zero. Physically, this condition implies that
all the {\it pure} Lorentz {\it scalar} (super)fields on the l.h.s. as
well as on the r.h.s. should {\it not} change at {\it all}. In other words, we have the following
\[
\tilde X^\mu\,[\tilde h^a(\xi,\theta, \bar\theta), \theta, \bar\theta] = {\cal X}^\mu\,[\xi, \theta, \bar\theta] \equiv  X^\mu(\xi),
\]
\[
\bar Q^\mu[\tilde h^a(\xi,\theta, \bar\theta)] = \bar Q^\mu(\xi), \qquad Q^\mu[\tilde h^a(\xi,\theta, \bar\theta)] = Q^\mu(\xi),  
\]
\[
T^\mu[\tilde h^a(\xi,\theta, \bar\theta)] = T^\mu(\xi), \eqno(D.7)
\]
so that we have the following super expansion:
\[
X^\mu(\xi, \theta, \bar\theta) = X^\mu(\xi) + \theta\,\bar Q^\mu(\xi) + \bar\theta\,Q^\mu(\xi) + \theta\,\bar\theta\,T^\mu(\xi)
\]
\[
\equiv X^\mu(\xi) + \theta\,(\bar s_B\,X^\mu) + \bar\theta\,(s_B\,X^\mu) + \theta\,\bar\theta\,(s_B\,\bar s_B\,X^\mu). \eqno(D.8) 
\]
It should be recalled that $s_B$ and $\bar s_B$ are the BRST and anti-BRST symmetry transformations that have been listed in Eqs. (6) and (9).
For the expansion of the type (D.9), Bonora-Tonin (BT) have found the mappings: $s_B \leftrightarrow \partial_{\bar\theta}|_{\theta = 0}, \,
\bar s_B \leftrightarrow \partial_{\theta}|_{\bar\theta = 0}$ in the context of BT-superfield approach to 4D non-Abelian 1-form 
(see, e.g. [21-23]) gauge theory 
(without any interaction with matter fields) where
the (anti-)BRST symmetry transformations as well as the CF-condition [24] have been found. We note that the HC 
[cf. Eq. (D.6)] implies, taking into account 
the inputs from (D.5), the following:
\[
Q^\mu = C^a\,\partial_a\,X^\mu, \qquad \bar Q^\mu = \bar C^a\,\partial_a\,X^\mu,
\] 
\[
T^\mu = C^a\,\partial_a\,\bar Q^\mu + \bar C^a\,C^b\,\partial_a\,
\partial_b\,X^\mu - \bar C^a\,\partial_a\,Q^\mu - k^a\,\partial_a\,X^\mu. \eqno(D.9) 
\]
The substitutions of the above into (D.8) and their interpretations in terms of the 
off-shell nilpotent (anti-)BRST symmetry transformations $(\bar s_B)s_B$
imply that:
\[
Q^\mu = s_B\,X^\mu = C^a\,\partial_a\,X^\mu, \qquad \bar Q^\mu = \bar C^a\,\partial_a\,X^\mu = \bar s_B\,X^\mu,
\]
\[
T^\mu = s_B\,\bar s_B\,(X^\mu) \equiv C^a\,\partial_a\,\big[\bar C^b\,\partial_b\,X^\mu\big] + C^a\,\bar C^b\,\partial_a\,\partial_b\,X^\mu
\]
\[
- \bar C^a\,\partial_a\,\big[C^b\,\partial_b\,X^\mu\big] - k^a\,\partial_a\,X^\mu
\]
\[
\equiv \big[C^a\,\partial_a\,\bar C^b - \bar C^a\,\partial_a\,C^b - k^b\big]\,(\partial_b\,X^\mu) - \bar C^a\,C^b\,\partial_a\,\partial_b\,X^\mu. \eqno(D.10)
\]
At this juncture, we demand the absolute anticommutativity (i.e. $\{s_B, \bar s_B\}\,X^\mu = 0$) of the (anti-)BRST
symmetry transformations $(\bar s_B)s_B $ which leads to the following:
\[
s_B\,\bar s_B\,X^\mu = s_B\,\big[\bar C^a\,\partial_a\,X^\mu\big] \equiv i \;
B^a\,\partial_a\,X^\mu - \bar C^b\,C^a\,\partial_a\,\partial_b\,X^\mu
\]
\[
- \; \bar C^a\,(\partial_a\,C^b)\,(\partial_b\,X^\mu),
\]
\[
-\,\bar s_B\,s_B\,X^\mu = - \bar s_B\,(C^a\,\partial_a\,X^\mu) \equiv - i\; \bar B^a\,\partial_a\,X^\mu - C^a\,\bar C^b\,\partial_a\,\partial_b\,X^\mu
\]
\[
+\; C^a\,(\partial_a\,\bar C^b)\,\partial_b\,X^\mu, \eqno(D.11)
\]
where we have taken the standard (anti-)BRST symmetry transfromations:
$s_B\,\bar C^a = i\,B^a, \, \bar s_B\,C^a = i\,\bar B^a,\, s_B\,B^a = 0, \, \bar s_B\,\bar B^a = 0$ and 
$s_B\,C^a = C^b\,\partial_b\,C^a, \,\bar s_B\,\bar C^a = \bar C^b\,\partial_b\,\bar C^a$. The last two
symmetry transformations have been obtained from the nilpotency requirements: $s_B^2 X^\mu = 0, \bar s_B^2 X^\mu = 0$, respectively. 
Each equation from (D.11) can be compared with the exact expression for
$T^\mu$ that has been obtained in Eq. (D.10). This comparison implies
\[
k^a = -\; i\,B^a + C^b\,\partial_b\,\bar C^a \equiv \,i\,\bar B^a - \bar C^b\,\partial_b\,C^a, \eqno(D.12)
\]
which leads to the derivation of the CF-type restriction within the framework of MBTSA. Thus, it is very interesting to note that
the requirement of the absolute anticommutativty property [cf. Eq. (D.11)] of the (anti-)BRST symmetry
transformations is responsible for the determination of $k^a$ [cf. Eq. (D.12)] which, ultimately, leads to
the derivation of the CF-type restrictions (11) within the ambit of MBTSA. \\

%It should be recalled that 
%the secondary field $k^a(\xi)$ is present as the coefficient of $\theta\bar\theta$ in the
%generalization of the {\it ordinary} 2D diffeomorphism to its counterpart on the (2, 2)-dimensional supermanifold.

\vskip 0.1cm

\noindent
{\bf Acknowledgements}\\

\noindent
Fruitful conversations with L. Bonora (SISSA, Trieste, Italy) are gratefully acknowledged 
as he has been a constant source of encouragement during the completion 
of this work. Thanks are also due to T. Bhanja, A. K. Rao and A. Tripathi for their critical
reading of the present manuscript. This work was presented at the $80^{th}$ birth-anniversary 
celebration  of Prof. G. Rajasekaran at Chennai Mathematical Institute, Chennai, during 
August 2016. This paper is dedicated to {\it him} because, for the first time, the author came to know
about the {\it basics} of string theory from {\it him} at the SERC School, I. I. Sc. Bengaluru (1985).\\

\noindent
{\bf Data Availability}\\

\noindent
No data were used to support this study.\\

\noindent
{\bf Conflicts of Interest}\\

\noindent
The authors declare that there are no conflicts of interest


\begin{thebibliography}{99}
\bibitem{GS}    M. B. Green, J. H. Schwarz, E. Witten, {\it Superstring Theory}, Vols. 1 and 2,\\
                Cambridge University Press, Cambridge (1987)
\bibitem{PL}    J. Polchinski, {\it String Theory}, Vols. 1 and 2,\\
                Cambridge University Press, Cambridge (1998)
\bibitem{LT}    D. Lust, S. Theisen, {\it Lectures in String Theory}, Springer-Verlag, New York (1989)
\bibitem{BBJ}   K. Becker,  M. Becker, J. H. Schwarz, {\it String Theory and M-Theory},\\ 
                Cambridge University Press, Cambridge (2007)
\bibitem{JS}    J. Scherk, {\it Rev. Mod. Phys.} {\bf 47}, 123 (1975) 
\bibitem{KW}    M. Kato, K. Ogawa, {\it Nucl. Phys. } B {\bf 212}, 443 (1983)    
\bibitem{LBNB}    L. Bonora, {\it  Nucl. Phys.} B  {\bf 912}, 103 (2016)
\bibitem{AP}    A. M. Polyakov, {\it Phys. Lett.} B {\bf 103}, 207 (1981)
\bibitem{AP1}   A. M. Polyakov, {\it Phys. Lett.} B {\bf 103}, 211 (1981)
\bibitem{KU}    T. Kugo, S. Uehara, {\it Nucl. Phys. } B {\bf 197}, 378 (1982)  
\bibitem{KO}    T. Kugo, I. Ojima, {\it Supple. Prog. Theor. Phys.}  {\bf 66}, 1 (1979) 
\bibitem{DLS}   D. Dudal, V. E. R. Lemes, M. S. Sarandy, S. P. Sorella, M. Picariello,\\
                {\it JHEP} {\bf 0212}, 008 (2002) 
\bibitem{DLSV}  D. Dudal, H. Verschelde, V. E. R. Lemes, M. S. Sarandy, S. P. Sorella, M. Picariello,
                A. Vicini, J. A. Gracey, {\it JHEP} {\bf 0306}, 003 (2003) 
\bibitem{RPMLB} L. Bonora, R. P. Malik, {\it Phys. Lett.} B {\bf 655}, 75 (2007)
\bibitem{RPM1}  L. Bonora, R. P. Malik, {\it J. Phys. A: Math. Theor.} {\bf 43}, 375403 (2010)
\bibitem{ATRPM}   A. Tripathi, A. K. Rao, R. P. Malik, in preparation 
\bibitem{AKR}   A. K. Rao, A. Tripathi, R. P. Malik, arXiv: 2012.12021 [hep-th]
\bibitem{BC}    B. Chauhan, A. K. Rao, A. Tripathi, R. P. Malik, arXiv: 1912.12909 [hep-th]
\bibitem{AT}    A. Tripthi, B. Chauhan, A. K. Rao, R. P. Malik, arXiv: 2010.02737 [hep-th] 
\bibitem{LBRPM3}   L. Bonora, R. P. Malik, in preparation
\bibitem{TLB}    L. Bonora, M. Tonin, {\it Phys. Lett. } B {\bf 98}, 48 (1981)
\bibitem{PLB1}   L. Bonora, P. Pasti, M. Tonin, {\it Nuovo Cimento} A {\bf 63}, 353 (1981)
\bibitem{LBP2}   L. Bonora, P. Pasti, M. Tonin, {\it Annals of Physics} {\bf 144}, 15 (1982)
\bibitem {CF}   G. Curci, R. Ferrari, {\it Phys. Lett.} B {\bf 63}, 91 (1976)
\end{thebibliography}
\end{document}